\documentclass[aps,prb,twocolumn]{revtex4}

\usepackage[dvips]{graphicx}
\usepackage{amssymb,amsfonts,amsmath}
\usepackage{eucal,bm}
\usepackage{gensymb,subfigure}
\usepackage{natbib}
\usepackage{booktabs}
\usepackage{times,color,url}

\begin{document}

\title{Hysteresis, Phase Transitions and Dangerous Transients in Electrical Power Distribution Systems}

\author{Charlie Duclut $^{a,b}$}
\author{Scott Backhaus $^{c,b}$}
\author{Michael Chertkov $^{d,b}$}

\affiliation{$^a$ ICFP, D\'epartement de Physique de l'ENS, 24 rue Lhomond, 75005 Paris, France}
\affiliation{$^b$ New Mexico Consortium, Los Alamos, NM 87544, USA}
\affiliation{$^c$ Materials, Physics \& Applications Division, Los Alamos National Laboratory, NM 87545, USA}
\affiliation{$^d$ Theoretical Division and Center for Nonlinear Studies, Los Alamos National Laboratory, NM 87545, USA}

\date{\today}

\begin{abstract}
The majority of dynamical studies in power systems focus on the high voltage transmission grids where models consider large generators interacting with crude aggregations of individual small loads. However, new phenomena have been observed indicating that the spatial distribution of collective, nonlinear contribution of these small loads in the low-voltage distribution grid is crucial to outcome of these dynamical transients.  To elucidate the phenomenon, we study the dynamics of voltage and power flows in a spatially-extended distribution feeder (circuit) connecting many asynchronous induction motors and discover that this relatively simple 1+1 (space+time) dimensional system exhibits a plethora of nontrivial spatio-temporal effects, some of which may be dangerous for power system stability. Long-range motor-motor interactions mediated by circuit voltage and electrical power flows result in coexistence and segregation of spatially-extended phases defined by individual motor states--a ``normal'' state  where the motors' mechanical (rotation) frequency is slightly smaller than the nominal frequency of the basic AC flows and a ``stalled'' state where the mechanical frequency is small.  Transitions between the two states can be initiated by a perturbation of the voltage or base frequency at the head of the distribution feeder. Such behavior is typical of {\it first-order phase transitions} in physics, and this 1+1 dimensional model shows many other properties of a first-order phase transition with the spatial distribution of the motors' mechanical frequency playing the role of the {\it order parameter}. In particular we observe (a) propagation of the phase-transition front with the constant speed (in very long feeders); and (b) hysteresis in transitions between the normal and stalled (or partially stalled) phases.

\end{abstract}


\maketitle

\underline{\bf Popular Summary:}
Large electrical generators interacting over national-scale electrical transmission grids constitute a well-studied dynamical system. Rarely discussed and poorly understood are the dynamics of neighborhood-scale distribution grids extending from transmission substations to the multitude of individual customers.  However, the changing nature of electrical loads, e.g. the increasing prevalence of induction motors in residential air conditioning units, is creating distribution-grid dynamical processes that, when excited by unremarkable transmission-grid disturbances, lead to irreversible transitions with major impact on the reliability of transmission grids. Here, we present a unique model and analysis of these dynamics that allow analogy with other physical systems enabling rapid progress by leveraging knowledge developed in physics.

In contrast to usual approaches, we develop a spatially-continuous model of distribution-grid dynamics to investigate the collective dynamics that arise when many induction motors, which are individually nonlinear and hysteretic (bi-stable), are coupled via power flows and voltage evolution within a distribution grid. Normal-size perturbations excite these collective dynamics initiating soliton-like fronts that travel though the distribution grids where passage of the fronts results in transitions of individual motors between a normal state and an undesirable stalled state.  Individual bi-stability of each motor promotes globally hysteric behavior that is reminiscent of well-known first-order phase transition dynamics found in other physical systems.

Important extensions of this physics-based understanding include the ability to model the dynamics of the billions active "smart grid" loads predicted to revolutionize the electrical power system.

\section{Introduction}
\label{sec:intro}

Power systems are used to generate and transfer energy to electrical loads. In today's grid, generation is primarily done at large, centralized power stations ($\sim$100's of MW) and the transfer primarily occurs via alternating currents (AC) in national-scale, highly-meshed, high-voltage transmission grids.  A subset of nodes (substations) in the transmission grid transform the high voltage to a medium voltage level and interface to distribution grids, however, another change occurs at the substation.  The meshed network of national-scale transmission changes to many radial or tree-like structures in the distribution system whose spatial extent is only $\sim$ 1-10 km.  Each radial circuit, also called a ``feeder'', distributes the power delivered to the substation by the transmission system to the thousands of small electrical loads ($\sim$ 1 kW) spatially spread along its length.

Even though AC electrical generation and transmission grids are extended over large spatial scales, they are synchronized, i.e. power flows over the transmission lines create dynamical coupling between the large rotating generators forcing them to rotating in unison.  Perturbations to this synchronized system results in dynamics and transients spanning a large range of temporal scales from milliseconds to many minutes.  Many studies have addressed the dynamics of these large-scale transmission grids. Although many unresolved dynamic problems in the transmission grid remain, recent years have witnessed new phenomena that have refocused our attention on dynamics and transients occurring in the smaller-scale {\it distribution} circuits\cite{08load_modeling,09FIDVR}.  Although these phenomena occur on smaller scales, they involve collective behavior of many individual small nonlinear loads, and even coupling several distribution circuits, creating a significant impact on the larger-scale transmission system.

Perhaps the most drastic of these phenomena is voltage collapse \cite{Weedy1968,Venikov1975,Venikov1977,Taylor1994,Cutsem1998,VanCutsem2000}. Here, a quasi-static increase in loading pushes the distribution feeder to a bifurcation where the stationary normal/high voltage solution is lost and the feeder ``collapses'' to an undesirable low voltage solution that is dangerous for power system stability.  However, even for moderately loaded feeders that are far from this critical point, the nonlinearity of electrical loads may result in the emergence of multiple stationary solutions. In contrast to the case just described, these solutions cannot be reached via quasi-static evolution of the electrical loads. Instead, a significant and nonlinear perturbation to the feeder creates a dynamical trajectory that terminates in one of these additional solutions that may also be a ``bad solution'' from the standpoint  of voltage level, power system losses, stability, and equipment damage. To understand the possibility of this unwelcome outcome one needs to go beyond the traditional static description and analyze dynamics of the distribution system.

The spatio-temporal dynamics we seek to describe occur within an individual radial distribution feeder that connects many ($\sim$ thousands) small loads to a substation. We consider {\it electro-mechanical} dynamics occurring on scales ranging from fractions of a second to tens of seconds and analyze the spatial distribution of power flows along the circuit and the spatio-temporal transients stimulated by exogenous disturbances in the voltage and base frequency at the head of the distribution feeder.  Such disturbances, which primarily originate from faults and/or irregularities in the high-voltage transmission system, propagate through from load to load via power flows in the distribution feeder.  The propagation is affected by the {\it electro-mechanical} response of individual loads, and here we focus on the effects of nonlinear loads such as asynchronous (i.e. induction) motors. Typical {\it electro-dynamic} transients propagate with speed comparable to the speed of light and damp out in tens of milliseconds and thus are not important in our analysis.  On the other hand, composition of loads connected to a distribution feeder changes on much longer time scales ($\sim$ minutes) and are taken as fixed on the time scale of {\it electro-mechanical} dynamics of interest in this work.

An interesting and key feature of the dynamics is the long-range coupling between the spatially distributed loads created by power flows along the feeder.  These coupled dynamics are nontrivial to model and investigate, however, they are also extremely important for practical power engineering because our model solutions reveal serious problems for control and operation of power systems.  One such problem that motivates our study is the phenomenon of the Fault-Induced Delayed Voltage Recovery (FIDVR) \cite{92WSD,97Sha,98PHH,08load_modeling,09FIDVR}.  A FIDVR event is typically initiated by a fault on the transmission grid near a substation creating large fault currents that temporarily depress the voltage at the substation, perhaps for as little at two cycles of the nominally 50/60 Hz AC frequency ($\sim$~30 msec). The voltage depression propagates into the substation's distribution feeders causing an almost instantaneous reduction in the {\it electrical} torque generated by the connected induction motors, however, the {\it mechanical} torque on the motors does not change instantaneously and the motors begin to decelerate.  If the transmission fault and voltage depression last long enough, many of the induction motors along the feeder may stall. When the fault on the transmission grid is cleared, the voltage at the substation returns to near normal levels, however, a stalled induction motor draws large in-rush currents while at near zero rotation speed (mechanical frequency).  The time synchronization of these in-rush currents cause large voltage drops in the distribution feeder and may hold the voltage at locations remote from the substation below a critical voltage for restarting.  Crucially, these remote motors remain stalled (near zero mechanical frequency), and their large current draw stabilizes this spatially-extended, partially stalled state.

The tendency for a transmission fault to result in FIDVR may depend on many fault, distribution feeder, and motor parameters, e.g: voltage drop magnitude and duration; length, resistance, and reactance of the feeder; type, rotational inertia, and density of the induction motor loading; and possible post-fault corrective control actions. The qualitative description of FIDVR given above provides an intuitive understanding of how some of these parameters affect the dynamics leading to the undesirable and potentially dangerous partially-stalled state.  It certainly indicates that, without some sort of corrective actions, FIDVR will become more and more frequent because of the recent trends to more air conditioning driven by easy-to-stall, low inertia motors.  However, the dynamics that lead to FIDVR are not generally understood and thus presumed somewhat mysterious in power engineering practice.  The goal of this manuscript is to provide understanding of these interesting and practically important distribution grid dynamics.

FIDVR is an example of a broader class of problems where physics and dynamical systems modeling can provide significant insight and predictive power that are generally lacking. Another example is given by electro-mechanical waves propagating through transmission grids \cite{Parashar2004,Thorp1998}, as well as transients associated with the loss of synchrony in power systems \cite{10DB,12DCB}. The key unifying feature of all these phenomena is in the nontrivial interplay of spatial coupling of individual (possibly nonlinear) dynamics via power flows over the electrical network. We believe that important insights into these complex dynamics can be gained by approaching such spatio-temporal phenomena from a homogenized prospective, i.e. studying the electrical grid not as a set of individual devices but rather as a spatially extended and continuous medium in the limit where the number of individual elements of the power system becomes infinite. This abstract continuous-medium approach, pioneered for the case of electro-mechanical waves over transmission systems in \cite{Parashar2004,Thorp1998} and for the case of a radial/linear distribution system in \cite{11CBTCL} is advantageous as it enables (a) a simpler analysis and deep qualitative physical understanding of the underlying phenomena (e.g. of FIDVR and electro-mechanical waves); (b) flexibility in simulations; and (c) developing model reduction algorithms for faster state estimation and system simulation.

Motivated by the discussion above, the main goal of this manuscript is to
formulate the simplest but still realistic 1+1 (space+time continuous) model that predicts and explains interesting and important spatio-temporal phenomena in a distribution feeder loaded with induction motors which can be in a normal or stalled state. The most important results reported in this manuscript are
\begin{itemize}
\item Extending the previous works \cite{98PHH,02PKMDUZ,10Les,08load_modeling,08LKU,10KD}, we show that if the local voltage falls sufficiently low, an individual asynchronous motor can be in either of the following two states: (1) a normal state characterized by mechanical frequency $\omega$ which is slightly lower than the base electrical frequency $\omega_0$ of the system; (2) a stalled state characterized by low or zero mechanical frequency. Both states are locally stable but can evolve into each other under sufficiently large perturbations.
\item In sufficiently long feeders,  a partially stalled phase can emerge where the feeder splits into head and tail parts with the motors of the head (tail) being in the normal (stalled) state. Motors in the stalled state may occupy the entire distribution feeder or, if the feeder is very long, stalled portion can co-exist with the normally running one. In the latter case, there exist multiple, partially stalled phases characterized by different proportions of the head (normal) and tail (stalled) parts.
\item The steady partially stalled phases can be interpreted as showing coexistence of the two states where the local mechanical frequency (of the motors) play the role of "order parameter".  Transitions between the phases are classified as first order, using standard physics terminology.
\item These transitions are hysteretic (not reversible), i.e. a perturbation leading to transition from the normal phase to the stalled phase is not an inverse of the perturbation leading from the stalled phase to the normal phase. The dynamics of the two transitions are also different, in particular fronts of the phase transitions have different shapes, and to stabilize one transition can take significantly longer than the other.
\end{itemize}

Material in the manuscript is organized as follows. Static models of single induction motor, two-bus system, and the DistFlow equations of \cite{89BWa,89BWb} are discussed in Sections \ref{subsec:IM},\ref{subsec:DIM} and \ref{subsec:DF}, respectively. A dynamic model of a distribution feeder loaded with induction motors and 1+1 space-time continuous model of this feeder are introduced and discussed in Sections \ref{subsec:CDF} and \ref{subsec:pde}. Section \ref{subsec:loc_gl} discusses how single-motor bi-stability translates into the emergence of multiple phases of the feeder with supporting numerical experiments in Section \ref{subsec:loc_gl_num} and special features of the phase transitions in Section \ref{subsec:special}. Section \ref{sec:FIDVR} is devoted to in-depth discussion of the results of our numerical experiments. Section \ref{subsec:fault} discusses the dynamics following a fault at the head of the feeder. Section \ref{subsec:recovery} analyzes the recovery from a stalled state.  Section \ref{subsec:dyn_trans} explores the phase space of parameters that governs whether or not a feeder will enter a stalled or normal state following a fault.  Finally, we summarize and describe a path forward in Section~\ref{sec:disc}. Auxiliary information explaining details of our simulations can be found in Appendix \ref{sec:methods}. Appendix \ref{sec:movies} provides captions for the illustrative movies of the phase transition and fault recovery processes available as a Supplementary Information (SI) to the manuscript.

\section{Single Induction Motor Models}
\label{sec:SIMM}

\subsection{Static Motor Model}
\label{subsec:IM}

\begin{figure}
	\centering
	\includegraphics[width=.9\linewidth]{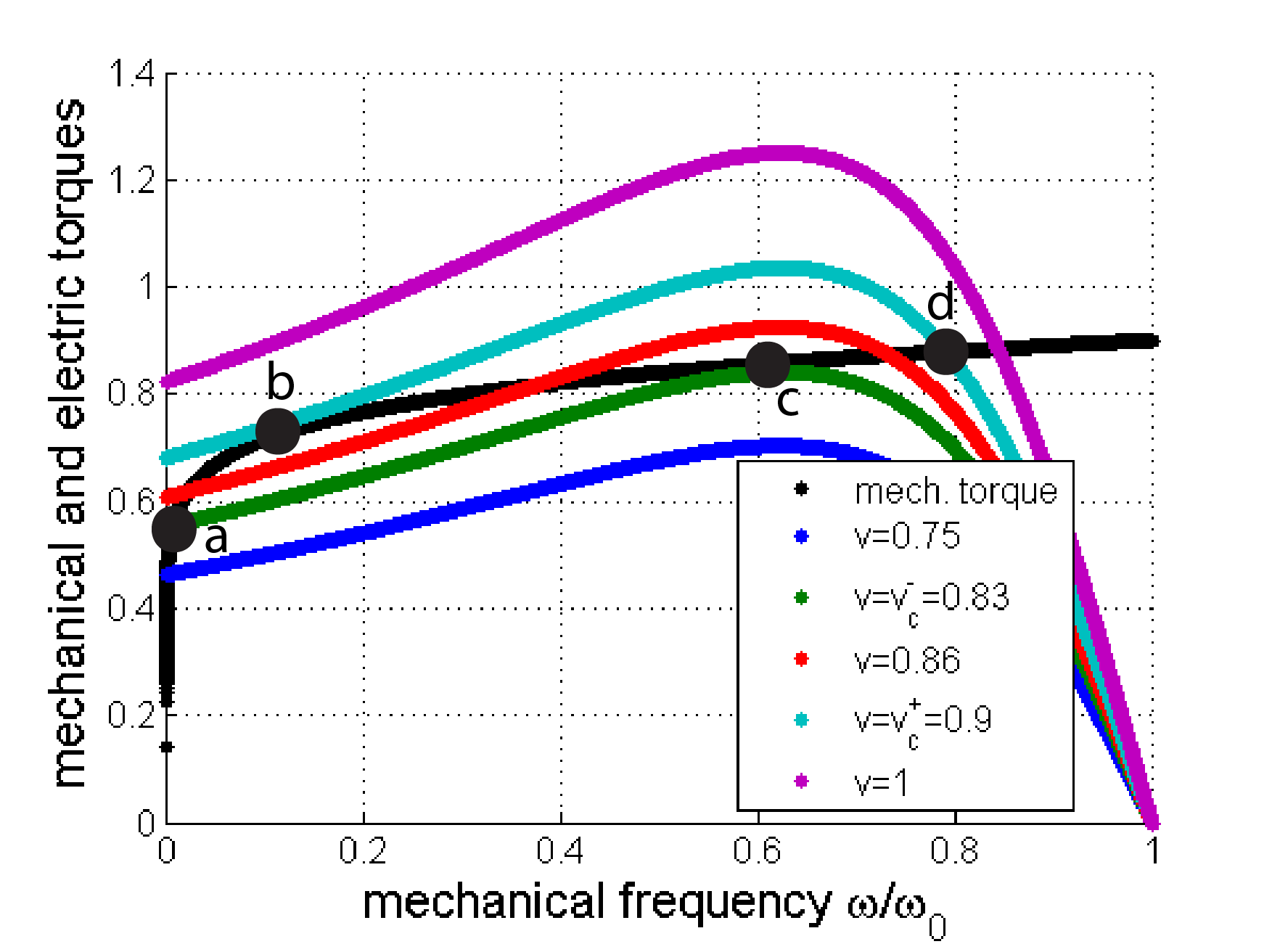}
	\caption{
Electric and mechanical torques as functions of the mechanical frequency $\omega/\omega_0$ for a range of motor terminal voltages $v$, reference mechanical torque $T_0=0.32$, and $\alpha=0.1$.  For the $v=0.86$ electrical torque curve (red), there are three equilibrium solutions indicated by intersections with the mechanical torque curve (black).  The solution with the highest $\omega/\omega_0$ is the ``normal'' stable solution with the induction motor rotating near the grid frequency $\omega_0$.  The ``stalled-state'' with $\omega/\omega_0\simeq 0$ is also stable while the intermediate solution is unstable.  For $v>v_c^+=0.9$ (light blue curve), there is only one solution corresponding to the normal state.  For $v<v_c^-=0.83$ (green curve), there is only one solution corresponding to the stalled state.  The points ($a,b,c,d$) correspond to the same labels in Fig.~\ref{fig:hysteresis-single}.}
	\label{fig:torques}
\end{figure}

Induction motors play a significant role in FIDVR, as it is currently understood. Here, we describe the features of inductions motors that are important for the rest of our work. Although motor dynamics will be added later, we first adopt a simple static electrical model of an induction (asynchronous) motor rotating at mechanical frequency $\omega$ and connected to a distribution circuit being driven at a base frequency $\omega_0,$\cite{98PHH}
\begin{eqnarray}
&& P=\frac{s R_m v^2}{R^2_m+s^2 X^2_m},
\label{p-motor}\\
&& Q=\frac{s^2 X_m v^2}{R^2_m+s^2 X^2_m}.
\label{q-motor-stat}
\end{eqnarray}
Here, $P$ and $Q$ are real and reactive powers drawn by the motor; $s=1-\omega/\omega_0$ is the slip parameter of the motor, $0\leq s<1$; $v$ is the voltage at the motor terminals; $X_m,R_m$ are internal reactance and resistance of the motor, usually $R_m/X_m =0.1\div 0.5$. For steady rotation frequency at $\omega$, the balance of electric and mechanical torques for the induction motor is
\begin{align}
& \frac{P}{\omega_0}=T(\omega/\omega_0),
\label{torque-balance}
\end{align}
where $T(\omega/\omega_0)$ is the rotation speed-dependent torque applied to the motor shaft by the mechanical load, which is typically parameterized by
\begin{eqnarray}
T(\omega/\omega_0)=T_0 \left(\frac{\omega}{\omega_0}\right)^\alpha.
\label{torque-model}
\end{eqnarray}
Here, $T_0$ is a reference mechanical torque and $\alpha$ is indicative of different types of mechanical loads with $\alpha=1$ typical of fan loads and $\alpha<1$ typical of air-conditioning loads. If $\alpha<\alpha_c \simeq 1$ and $T_0$ is fixed, one observes the emergence of three solutions when $v$ is in a range between two spinodal voltages $v_c^-$ and $v_c^+$, i.e. for $v_c^-<v<v_c^+$ (see Fig.~\ref{fig:torques}). These solutions have widely different $\omega$ which leads to hysteresis and the interesting dynamical behavior explored in the rest of this  manuscript.

\begin{figure}
	\centering
	\includegraphics[width=.9\linewidth]{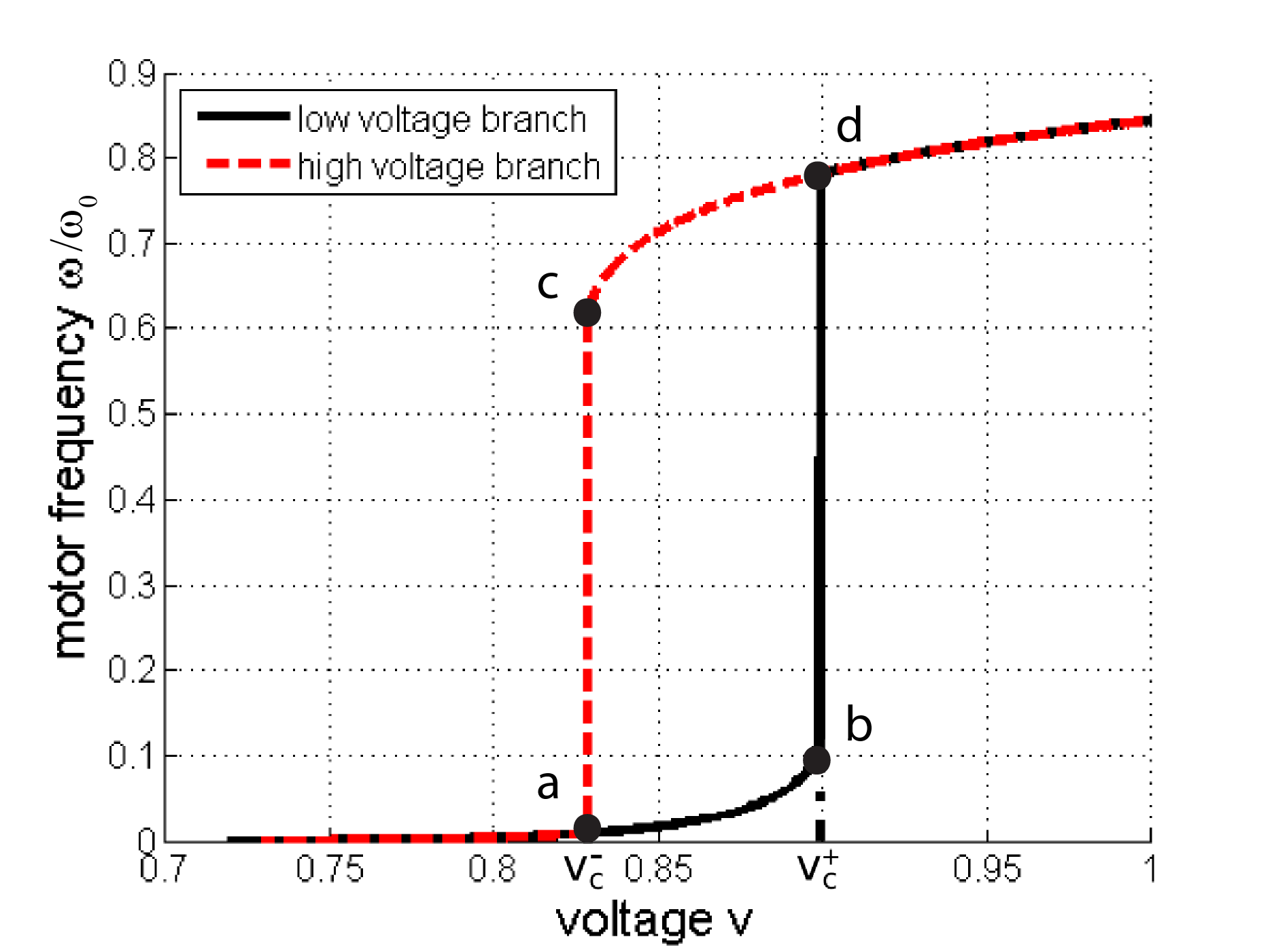}
	\caption{Hysteretic behavior of an induction motor in Fig.~\ref{fig:torques} as the voltage $v$ at its terminals is varied.  The dashed red (solid black) curves indicate the path of equilibrium states as the voltage $v$ is decreased (increased) starting from the high-voltage normal (low-voltage stalled) state.  The vertical lines at the spinodal voltages $v_c^\pm$ indicated the abrupt hysteretic transitions between states.  $v_c^\pm$ correspond to the same labels in the legend of Fig.~\ref{fig:torques} and in Fig.~\ref{fig:pq-motor}.  The points ($a,b,c,d$) correspond to the states where the motor must make transitions from normal to stalled ($c\rightarrow a$) and from stalled to normal $(b \rightarrow d)$.}
	\label{fig:hysteresis-single}
\end{figure}

Stability analysis (see Section \ref{subsec:DIM} for details) shows that the two extreme solutions ($\omega\approx 0$ and $\omega \approx\omega_0$) are both stable while the solution in the middle is unstable. The consequence is hysteretic behavior of the motor frequency $\omega$ as a function of the voltage $v$, as displayed in Fig.~\ref{fig:hysteresis-single}.  Starting  in the high-voltage normal state (say $v\sim 1$), we decrease $v$ slowly along the dashed red curve passing through state $d$. If we further decrease $v$ to state $c$, the normal state suddenly disappears, and the motor makes a transition to the stalled state at $a$.  Similarly, if we start from the low-voltage stalled state (say $v\sim 0.75$ on the black curve) and $v$ is increased slowly through state $a$ to $b$, the stalled state disappears and the motor makes a transition to the normal state at $d$.  The states ($a,b,c,d$) are also marked in Fig.~\ref{fig:torques}, and the same hysteresis loop can be traced out there.

\begin{figure}
\centering
\includegraphics[width=0.45\textwidth]{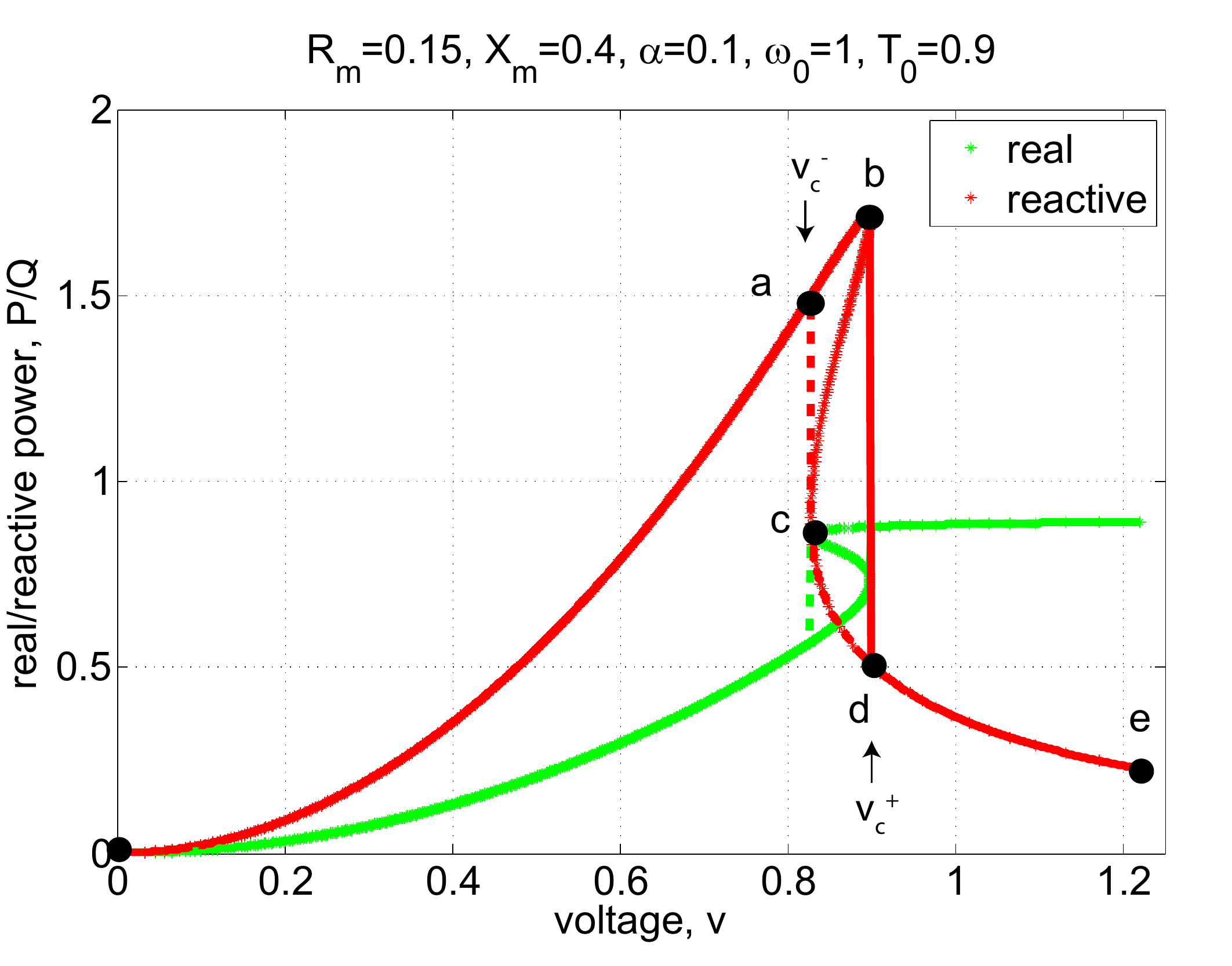}
\caption{Typical ($P,Q$) versus $v$ curves for the same motor as in Fig.~\ref{fig:torques} and \ref{fig:hysteresis-single} and described by Eqs.~(\ref{p-motor},\ref{q-motor-stat},\ref{torque-balance},\ref{torque-model}). Three solutions (two stable and one unstable) are observed between the spinodal voltages, i.e. $v_c^-<v<v_c^+$.  Solid and dashed curves (the latter partially covered by solid) show trajectory of the system under adiabatic evolution starting from low-voltage and high-voltage regimes respectively. The points ($a,b,c,d$) label the reactive power curve (red) and correspond to the same labels in Fig.~\ref{fig:torques} and Fig.~\ref{fig:hysteresis-single}. Here, $X_m/R_m=0.375$.}
\label{fig:pq-motor}
\end{figure}

After elimination of the auxiliary variable $\omega$, Eqs.~(\ref{p-motor},\ref{q-motor-stat},\ref{torque-balance},\ref{torque-model}) describe the dependence of the power flows $(P,Q)$ on terminal-voltage. If the reactance-to-resistance ratio of the motor, $X_m/R_m$, is sufficiently small, the hysteresis observed for mechanical frequency in Fig.~\ref{fig:hysteresis-single} translates into hysteresis of real and reactive powers as seen by the multi-valued dependence of $(P,Q)$ on $v$ in Fig.~\ref{fig:pq-motor}.  Between the spinodal voltages $v_c^-$ and $v_c^+$, there exists three solutions with different values of $(P,Q)$ for the same value of voltage.  Of the three solutions, the top and bottom are stable while the middle solution is unstable. Similar to Fig.~\ref{fig:hysteresis-single}, we can follow an adiabatic evolution of the motor terminal voltage.  Following the reactive power curve (red) and starting in the normal state with $v\sim 1$, we decrease the voltage through state $d$ and to state $c$.  Any further reduction of $v$ forces the motor to make a discontinuous jump to state $a$ which is accompanied by a large increase in reactive power $Q$.  Alternatively, we may start in the stalled state with $v\sim 0.7$ and slowly increase $v$ through state $a$ to $b$ where the motor is forced to jump to state $d$ accompanied by a discontinuous decrease in $Q$.  As discussed later, these discontinuous jumps in $Q$ play a significant role is stabilizing the spatially extended stalled state and in the recovery from the stalled state to normal state.



\subsection{Dynamical Motor Model}
\label{subsec:DIM}
To study the important yet generic aspects of the dynamics in distribution circuits, we generalize Eqs.~(\ref{p-motor}-\ref{torque-balance}) to include induction motor dynamics.  
Considering for the moment a single induction motor, an imbalance in electrical and mechanical torques will cause a change in the motor's rotational frequency given by
\begin{eqnarray}
M\frac{d}{dt}\omega=\frac{P}{\omega_0}-T_0\left(\frac{\omega}{\omega_0}\right)^\alpha,
\label{omega-relaxation}
\end{eqnarray}
where $M$ is the motor's moment of inertia.  Torque imbalances in Eq.~(\ref{omega-relaxation}) can be driven in two ways: directly via changes in the base frequency $\omega_0$  in Eq.~(\ref{omega-relaxation}) or indirectly via changes $P$ driven by changes in $v$ in Eq.~(\ref{p-motor}).  The coupling of Eq.~(\ref{omega-relaxation}) to $\omega_0$ will not be strong because $\omega_0$ is determined by the {\it global} balance of generation and load across the entire transmission system.  We would not expect the local distribution dynamics under consideration here to affect $\omega_0$ to a degree that we would have to consider its effect back on the distribution dynamics via Eq.~(\ref{omega-relaxation}).  Therefore, we can generally ignore the dynamics of $\omega_0$ and consider it an imposed exogenous parameter.

In contrast, changes in local voltage are strongly coupled to changes in the local flow of real and reactive power.  As we have seen in Sec.~\ref{subsec:IM}, changes in voltage can lead to drastic and hysteretic changes in a motor's frequency $\omega$ resulting in a strong coupling back to the dynamics in Eq.~(\ref{omega-relaxation}).  Therefore, we must consider the possibility that voltage dynamics are important for distribution feeder dynamics.  However, the dynamics of $v$ are fundamentally different than $\omega$ because the relaxation of $v$ is entirely electrical as opposed to the mechanical dynamics of $\omega$, i.e.
\begin{align}
Q=\frac{s^2 X_m}{R^2_m+s^2 X^2_m}\left(v^2+(\tau/2) \frac{d}{dt}{v^2}\right), \label{q-motor}
\end{align}
where $\tau$ is the characteristic time of this purely electrical process (See e.g. the Appendix of Pereira {\it et al}\cite{02PKMDUZ}). In our numerical experiments, we observe that the important basic phenomena discussed in the manuscript are much more sensitive to variations in the moment of inertia $M$ than to variations in $\tau$ and that setting $\tau=0$ still reveals the dynamical processes important for understanding FIDVR.  With $\tau=0$, Eq.~(\ref{q-motor}) is now equivalent to its static version in Eq.~(\ref{q-motor-stat}).


With the dynamics now fully specified by Eq.~(\ref{omega-relaxation}), we can now justify the local (single motor) stability claims made in Section \ref{subsec:IM} by simple inspection of the equilibrium states in Fig.~\ref{fig:torques}. Here stability/instability is understood in terms of the temporal decay/growth of small perturbations to the dynamics of Eq.~(\ref{omega-relaxation}).  The black curve in this Fig.~\ref{fig:torques} represents the mechanical torque on the motor while the colored curves are the electrical torques (each representing a different  $v$).  Consider state $d$ in Fig.~\ref{fig:torques} which is representative of the normal states with $\omega/\omega_0\sim 1$.  If the motor speeds up slightly (moves to the right along the light blue curve), the mechanical torque becomes larger than the electrical torque and the motor decelerates returning to $d$.  If the motor slows slightly (moves left), the electrical torque becomes larger while the mechanical torque decreases returning the motor to state $d$.  All of the normal states, i.e. those like $d$ with $\omega/\omega_0\sim 1$, have similar behavior and are therefore stable.  State $a$ in Fig.~\ref{fig:torques} is representative of the stalled states, and following the same logic, we find that the mechanical torque is larger for a higher $\omega$ (and smaller for a lower $\omega$) showing that all of the stalled states are stable.  Following the same logic, we find the the states in between the normal and stalled states (e.g. the state given the intersection of red and black curves in Fig.~\ref{fig:torques}) are unstable.


\section{Spatially-Continuous Feeder Power Flow Model}

In Section~\ref{sec:SIMM}, we described the dynamics of isolated induction motors, i.e. motors whose terminal voltage $v$ is specified and not determined in part by interactions with other induction motors or electrical loads.  In this Section, we consider power flow models responsible for creating the  long-range coupling between the individual, local induction motor dynamics.  We start with a well-known discrete power flow model which we then homogenize into a spatially continuous ODE representation.  We then incorporate a homogenized version of the individual induction motor dynamics to create a PDE representation of electrical feeder dynamics.

\subsection{Discrete Power Flow Model--Dist Flow Equations}
\label{subsec:DF}

\begin{figure}
	\centering
	\includegraphics[width=.75\linewidth]{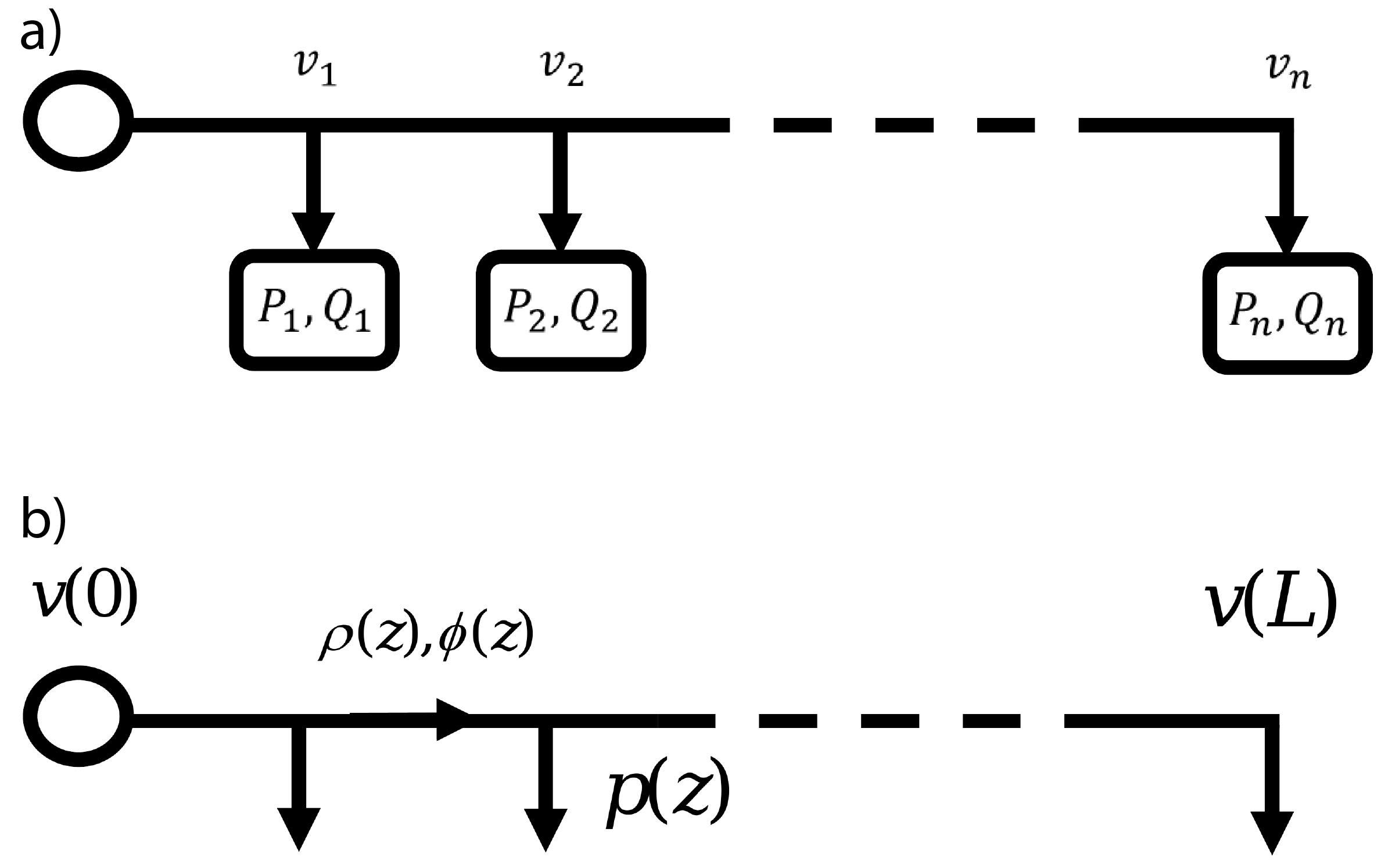}
	\caption{A feeder line modeled with a) a discrete set of electrical loads and with b) a continuous distribution of loads.}
	\label{fig:continuous-model}
\end{figure}

The flow of electric power in the quasi-static approximation is controlled by the Kirchoff laws. The DistFlow equations \cite{89BWa,89BWb} are these equations, written in terms of power flows and in a convenient form for the radial or tree-like distribution circuit with a discrete set of loads shown schematically in Fig.~\ref{fig:continuous-model}a,
\begin{eqnarray}
\rho_{n+1}-\rho_n & = & P_n-r_n \frac{\rho_n^2+\phi_n^2}{v_n^2},
\label{DF1}\\
\phi_{n+1}-\phi_n & = & Q_n-x_n\frac{\rho_n^2+\phi_n^2}{v_n^2},
\label{DF2}\\
 v_{n+1}^2-v_{n}^2 &=&-2 (r_n\rho_n+x_n \phi_n)\nonumber\\ &&-(r^2_n+x^2_n)\frac{\rho_n^2+\phi_n^2}{v_n^2}.
\label{DF3}
\end{eqnarray}
Here, $n=0,\cdots, N-1$ enumerates the sequentially-connected buses of the circuit, and $\rho_n,\phi_n$ are the real and reactive power flowing from bus $n$ to $n+1$. $v_n$ is the bus voltage, while $P_n$ and $Q_n$ are the overall consumption of real and reactive powers by the discrete load at bus $n$. The values of $r_n$ and $x_n$ are the resistance and reactance of the discrete line element connecting $n$ and $n+1$ buses.   The voltage $v_0$ at the beginning of the line is nominally fixed by control equipment, and there can be no flow of real or reactive power out of the end of the circuit.  These two observations provide the following boundary conditions:
\begin{eqnarray}
v_0 = v_0,\quad \rho_{N+1} = \phi_{N+1} = 0.
\label{DF-BC}
\end{eqnarray}
Eqs.~(\ref{DF1},\ref{DF2},\ref{DF3},\ref{DF-BC}) combined with the given real and reactive consumption pattern, $P_n,Q_n$ for $n=1,\cdots,N$, uniquely define profile of voltage, $v_n$, and power flows, $\rho_n,\phi_n$, along the circuit.

We note that the dynamical relaxation of distribution circuit power flows $\rho_n$ and $\phi_n$ will also occur on electrical time scales, i.e. much faster than the mechanical dynamics in Eq.~(\ref{omega-relaxation}). Therefore, the quasi-static power flows described in the DistFlow formulation\cite{89BWa,89BWb,10TSBCa,10TSBCb,11TSBC} in Eqs.~(\ref{DF1},\ref{DF2},\ref{DF3}) is a sufficient starting point for the phenomena discussed in the manuscript.

\subsection{Continuous Power Flow Model}
\label{subsec:CDF}

 Following \cite{11CBTCL}, the continuous form of the DistFlow equations is
\begin{eqnarray}
&& \partial_z \rho=-p-r\frac{\rho^2+\phi^2}{v^2},\label{p_cont}\\
&& \partial_z \phi=-q-x\frac{\rho^2+\phi^2}{v^2}, \label{q_cont}
\end{eqnarray}
where $z$ is the coordinate along the distribution circuit, $r$, $x$ are the per-unit-length resistance and reactance densities of the lines (assumed independent of $z$) and $p(z)$ and $q(z)$ are the local densities of real and reactive powers consumed by the density of the spatially continuous distribution of motors \cite{11CBTCL} at the position $z\in[0;L]$.  The power flows $\rho$ and $\phi$ are related to the voltage at the same position according to \cite{11CBTCL}
\begin{eqnarray}
\partial_z v=-\frac{r \rho+ x \phi}{v}. \label{v_eq}
\end{eqnarray}

\subsection{PDE Model of Feeder Dynamics}
\label{subsec:pde}

Instead of the standard voltage-independent $(p,q)$ model of distributed loads, discussed in \cite{11CBTCL}, we consider the more complex dynamical loads described above. The load densities $p(z)$ and $q(z)$ are related to $\omega(z)$ and $v(z)$ through the density versions of Eqs.~(\ref{p-motor},\ref{omega-relaxation},\ref{q-motor})
\begin{eqnarray}
\mu\frac{d}{dt}\omega&=&\frac{p}{\omega_0}-t_0\left(\frac{\omega}{\omega_0}\right)^\alpha,
\label{torque-cont}\\
p&=&\frac{s r_m v^2}{r^2_m+s^2 x^2_m},
\label{p-motor-cont}\\
q&=&\frac{s^2 x_m}{r^2_m+s^2 x^2_m}v^2. 
\label{q-motor-cont}
\end{eqnarray}
where the conversion to continuous form consists of replacing $X_m,R_m$ and $P,Q,T_0,M$ by the respective densities $x_m$, $r_m$ and $p$, $q$, $t_0$, and $\mu$.  The new boundary conditions are
\begin{eqnarray}
v(0)=v_0,\quad \rho(L)=\phi(L)=0.
\label{bc}
\end{eqnarray}

Equations~(\ref{p_cont}-\ref{bc}) form our PDE model of a distribution feeder loaded with induction motors.  We assume that the distributions of all the density parameters along the circuit are known, and in this initial work, we assume these densities are constant. Note that evolution in the model occurs solely due to temporal derivatives in Eq.~(\ref{torque-cont}) representing mechanical relaxation of the spatial distribution of motors.

\section{Phase Transitions and Hysteresis in a Feeder}
\label{sec:PT}

In this Section, we discuss the physical picture of phase transitions and hysteresis that emerges from analysis and simulations of the 1+1 space-time PDE model of Eqs.~(\ref{p_cont}-\ref{bc}). In Section~\ref{subsec:loc_gl}, we begin with a qualitative description of the hysteretic, phase transition-like behavior of the distribution feeder. Section \ref{subsec:loc_gl_num} discusses
numerical results in the framework of the qualitative arguments of Section~\ref{subsec:loc_gl}. Then, in Section~\ref{subsec:special} we regress again to provide general physics discussion of the phase transition special features observed in the simulations.

\subsection{From Local (Motor) to Global (Feeder): Qualitative Picture}
\label{subsec:loc_gl}

\begin{figure}
    \centering
    \includegraphics[width=0.9\linewidth]{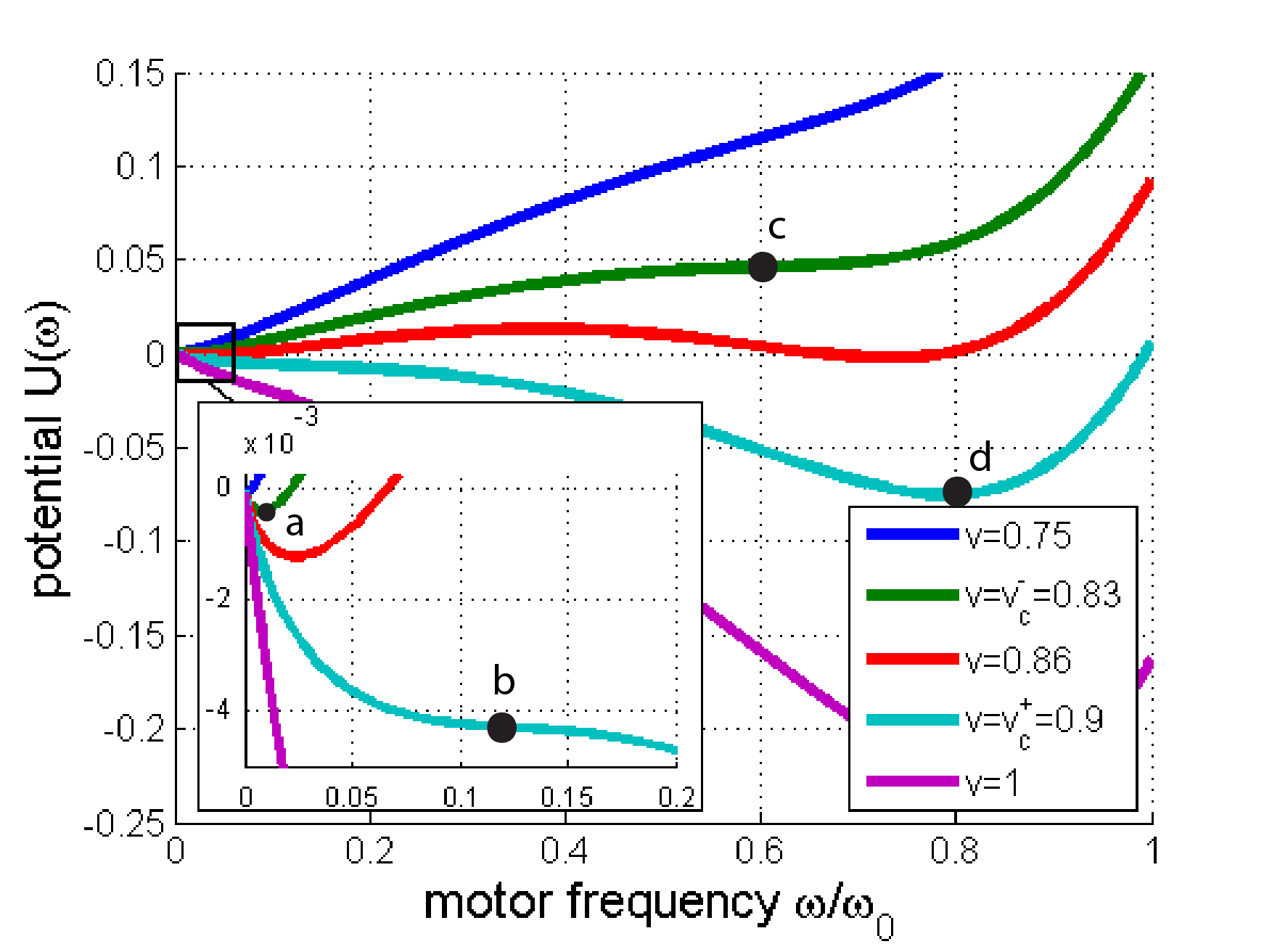}
    \caption{Local potential $U(\omega)$ for different values of the voltage $v$ and $\alpha=0.1$ demonstrating the continuous evolution of the normal and stalled states from only the normal state at high voltage (purple), coexisting normal and stalled states at intermediate voltages (red), and only the stalled state at low voltage (dark blue).  The light blue curve is for the spinodal voltage $v_c^+$ -- the boundary between the high and intermediate voltage cases. The green curve is for spinodal voltage $v_c^-$ -- the boundary between the low and intermediate voltage cases. The inset is a view of the region near $\omega=0 $ showing the emergence of a stalled $\omega\approx 0$ minimum of the potential for $v<v_c^+$.  The states labelled ($a,b,c,d$) are the same as those in Figs.~\ref{fig:torques}-\ref{fig:pq-motor}}
    \label{fig:landau}
\end{figure}

From the qualitative description of distribution circuit and induction motor dynamics and FIDVR events given in Section \ref{sec:intro}, we make an analogy between FIDVR and a first-order phase transition
\cite{79PP} where the $z$-dependent motor frequency $\omega(z)$ plays the role of the order parameter, i.e. $\omega \simeq \omega_0$ in the ``normal phase'' and $\omega\approx 0$ in the ``stalled phase''.  Recasting Eq.~(\ref{torque-cont}) in terms of an $\omega$-dependent potential $U$, we find
\begin{align}
  & \frac{\partial \omega}{\partial t} = - \frac{\partial U}{\partial \omega}, \label{eq:energy_dynamics}\\
  \text{where} \quad & U=\frac{1}{\mu} \int \limits_0^{\omega} \left(t_0\left(\frac{\omega^\prime}{\omega_0}\right)^{\alpha} - \frac{p(\omega^\prime;v)}{\omega_0} \right) \, d\omega^\prime,
  \label{energy}
\end{align}
and $p(\omega;v)$ is defined by Eq.~(\ref{p-motor-cont}).  The resulting effective potential $U(\omega;v)$ is shown graphically in Fig.~\ref{fig:landau} for different values of $v$.

We identify three different regimes in Fig.~\ref{fig:landau}: (a) high voltage $v>v_c^+$ (purple curve)  where the normal state is the only stable solution, (b) intermediate voltage  $v_c^-<v<v_c^+$ (red curve) where the normal state and the stalled state coexist, i.e. motors can be in either of the two states, and (c) low voltage  $v<v_c^-$ (dark blue curve) where the stalled state is the only stable solution.  The boundaries between these regions occur at the spinodal voltages $v_c^+$ (light blue curve) where the state jumps from $c$ to $a$ in Figs.~\ref{fig:torques}-\ref{fig:pq-motor} and at $v_c^-$ (green curve) where the state jumps from $b$ to $d$ in Figs.~\ref{fig:torques}-\ref{fig:pq-motor}

The high-voltage case (a) where the motors can only be in the normal state is the desired regime for electrical grid operations. The low-voltage case (c) is undesirable and is a result of the electrical torque at low voltage not being able to overcome the mechanical torque with the subsequent decline of $\omega$ to very small values.  Case (b) also allows for the motors to be in the normal state, but the existence of two minima in the intermediate voltage range $v_c^-<v<v_c^+$ (see Fig.~\ref{fig:landau}) is associated with the overlapping high and low voltage states in Fig.~\ref{fig:hysteresis-single} which can lead to {\it local} hysteretic behavior.  For example, a small voltage perturbation can kick a motor from the normal state over the potential barrier (see red curve in Fig.~\ref{fig:landau}) where Eq.~(\ref{eq:energy_dynamics}) shows that it subsequently relaxes to the stalled state.  A simple reversal of the perturbation does not necessarily lead to the reverse transition.  This entirely {\it local} hysteresis is important because it defines the possible states of motor operation, however, it is the long range interactions between the local motor behavior that creates phase transition-like behavior with a phase boundary between separate normal and stalled phases.

\begin{figure*}[t]
    \centering
    \subfigure[]{\includegraphics[width=0.33\textwidth]{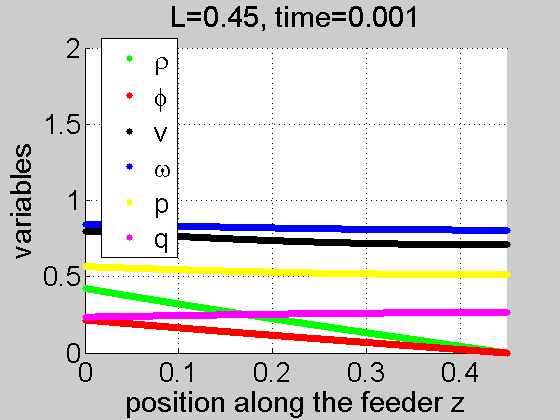}}
    \subfigure[]{\includegraphics[width=0.33\textwidth]{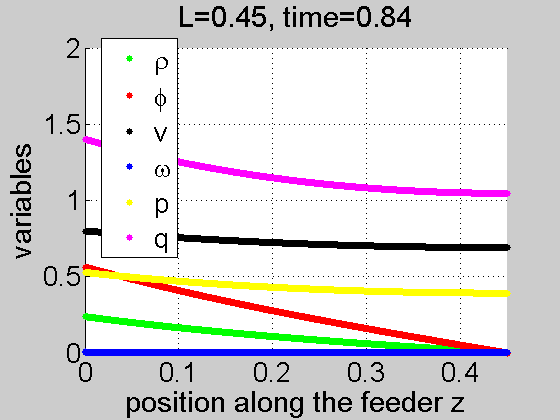}}
    \subfigure[]{\includegraphics[width=0.33\textwidth]{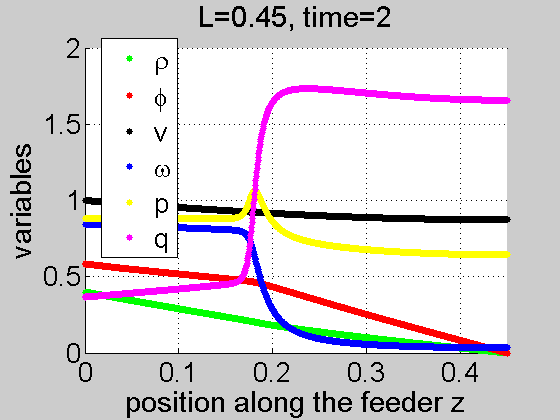}}
    \caption{Sequence of snapshots of a simulation for a feeder with $L=0.45<L_c$ that undergoes a short voltage fault. In this case $L$ is short enough so that the long-range interactions via power flows and voltage are insufficient to promote the local hysteresis at each motor into globally hysteretic behavior.  Snapshot (a) shows the beginning of the sequence (immediately post fault) where the small inertia of the motors maintains $\omega/\omega_0 \sim 1$ (blue) so that the the reactive power load density $q$ (pink) and reactive power flow $\phi$ (red) are low and the voltage profile (black) relatively flat, although near 0.8.
    Snapshot (b) shows the feeder at a time after the reduction of electrical torque has decelerated all of  motors to $\omega/\omega_0 \sim 0$ (blue) with an associated rise in reactive load density $q$ (pink) and reactive power flow $\phi$ (red) causing the voltage profile (black) to droop more than in (a).
    Snapshot (c) shows the situation after the fault is cleared with $v_0$ restored to 1.0.  Although the reactive power flow $\phi$ is still high, it is insufficient to force $v<v_c^+$  and the motors relax back to the normal state as the phase front propagates from left to right. The dynamical nature of this transition is evident from the real power load density $p$ (yellow).  The peak in $p$ at the front is a result of the acceleration of the motors as they transition from the stalled to normal state (see Eq.~(\ref{torque-cont})).
    The motors shown accelerating in (c) eventually reach $\omega/\omega_0 \sim 1$, and the feeder relaxes to a globally-normal state.
    See Appendix \ref{sec:movies} and Supplementary Information for respective movie (Movie 4: movie\_recovery.pdf).}
    \label{fig:no_hysteresis}
\end{figure*}

When a motor undergoes the local transition from a normal to a stalled state, its rotational frequency $\omega$ changes significantly.  Motors distributed along the circuit do not interact via $\omega$, however, changes in $\omega$ drive large changes in local reactive power density $q$ (see Fig.~\ref{fig:pq-motor}) which couples to all of the other motors via the power flows ($\rho,\phi$) and voltages in Eqs.~(\ref{p_cont}-\ref{v_eq}).  Crucially, if a perturbation causes a group of motors in the tail segment of the feeder ($z\sim L$) to enter the stalled states ($\omega\approx 0$ or $s\approx 1$), the increase in the local reactive load density $q$ drives an increase in the power flow $\phi$ all along the feeder, and Eq.~(\ref{v_eq}) shows that $v$ will be depressed at all $z$ along the feeder.  The voltage depression will be the largest at the tail ($z\sim L$), and if the depression is severe enough, the terminal voltage of the normal-state motors neighboring the stalled tail section will drop below $v_c^-$ and the local potential $U(\omega)$ changes from the purple or light blue curves of Fig.~\ref{fig:landau} to the green or dark blue curves.  Equation~(\ref{eq:energy_dynamics}) shows that these motors relax into the stalled state, further increasing the power flows $\phi$ everywhere along feeder. This phase transition front continues to propagate toward $z=0$ as the increases in local $q$ drive increases in the non-local $\phi$ which depress the non-local $v$. The voltage boundary condition at the head of the feeder (Eq.~(\ref{bc})) may stop the the front before its reaches all the way back to $z=0$, however, the establishment of a globally stalled or partially stalled phase is hysteretic because simply reversing the original spatially-local perturbation, i.e. returning the small section of motors in the tail to the running state, will not be able to overcome the globally-stalled state once it has become established.

\subsection{From Local to Global: Numerical Experiments}
\label{subsec:loc_gl_num}

\begin{figure}
    \centering
    \includegraphics[width=0.45\textwidth]{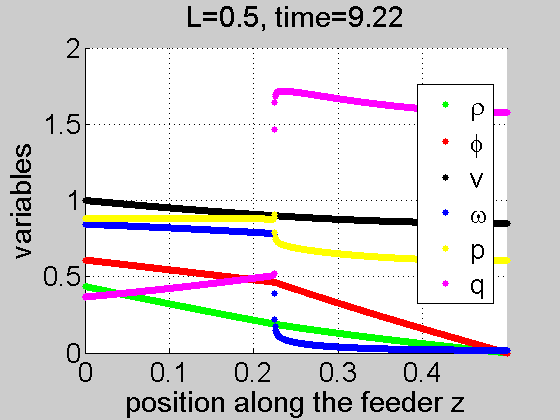}
    \caption{Voltage fault and subsequent hysteretic recovery for $L=0.50>L_c$. The sequence of events (fault, transient and initial recovery) are identical to the one shown in Fig.~\ref{fig:no_hysteresis}, however the recovery is not complete.  The additional loading from the motors between $z=0.45$ and $z=0.50$ generate long-range interactions that force $v<v_c^+$ for $z>0.22$.  Although $v_0$ has been restored to 1.0, the motors with $v<v_c^+$ do not recover to a normal state, even though many of them do have such a stable state.  In this case, the long-range interactions are are sufficiently strong to promote the local hysteretic behavior of each motor into globally hysteretic behavior.  See Appendix \ref{sec:movies} and Supplementary Information for respective movie (Movie 3: movie\_hysteresis.pdf).
    \label{fig:hysteresis}
    }
\end{figure}

Next, we perform numerical simulations of Eqs.~(\ref{p_cont}-\ref{bc}) to explore the qualitative dynamical description given above.  We examine how one property of the feeder, i.e. its length $L$, affects the qualitative description given above by performing two identical simulations except $L=0.45$ in Fig.~\ref{fig:no_hysteresis} and is only slightly longer at $L=0.5$ in Fig.~\ref{fig:hysteresis}.  Although the change in length is relatively minor, the final states of the two feeders are radically different.  In  both simulations, we start with all the induction motors (the only type of load considered here) in the normal state, i.e. $\omega\sim \omega_0$.  At $t=0$, a perturbation is applied where $v_0$ is abruptly lowered to 0.8 and held there long enough so that all of the motors stall.  Subsequently, $v_0$ is restored to 1.0 and the evolution of the motors and feeder variables are monitored.  The forced evolution of $v_0$ emulates the behavior that would be driven by a nearby fault on the transmission system supplying the feeder and its substation.

Figure~\ref{fig:no_hysteresis}a shows the state of the $L=0.45$ feeder immediately after the fault is applied.  The voltage (black curve) starts at $v(0)=0.8$ and droops only slightly.  Although $v<v_c^-$, the inertia of the motors (although small) keeps them temporarily at $\omega/\omega_0\sim 1$ (blue curve) so that their local reactive loads $q$ (pink curve) remain low as does the non-local reactive power flow (red curve).  However, because $v<v_c^-$ the motors no longer have an equilibrium state at $\omega/\omega_0\sim 1$ (see dark blue curve in Fig.~\ref{fig:landau}) and Eq.~(\ref{eq:energy_dynamics}) forces them to relax to the stalled state which is evident in Fig.~\ref{fig:no_hysteresis}b where $\omega/\omega_0\sim 0$ (blue curve) all along the feeder.  All of the induction motors have now made the transition to the upper branch of the reactive power curve in Fig.~\ref{fig:pq-motor} near to state $a$ which is reflected by the increase in $q$ (pink curve) and the reactive power flow $\phi$ (red curve).  Subsequent to Fig.~\ref{fig:no_hysteresis}b, $v_0$ is restored to 1.0, and a left-to-right propagating phase front is formed (see blue curve in Fig.~\ref{fig:no_hysteresis}c) where the normal and stalled phases are segregated with all the motors to the left of the front in the normal state while those to the right are in the stalled state.  The motion of the finite-thickness phase front is evident from the local real power load density (yellow curve in Fig.~\ref{fig:no_hysteresis}c).  The peak in $p$  (above the steady-state values observed far to the left or right of the front) is responsible for the acceleration of the motors within the front, and as the motors in front accelerate to $\omega/\omega_0\sim 1$, the front advances to the right.  A left-propagating front would show a downward peak in $p$.   After $v_0$ has been restored to 1.0, the reactive power $q$ drawn by the stalled motors along this feeder is insufficient to create a large enough $\phi$ to lower $v$ into the intermediate range $v_c^-<v<v_c^+$, and the feeder completely recovers to its initial state, i.e. the reversal of the initial perturbation restores the feeder state.  Alternatively, the implicit long-range power flow interactions, expressed in a globally depressed voltage profile, are insufficient to turn the local hysteresis into global hysteresis. The dynamic version of this simulation is provided in the SI (movie 4: movie\_recovery.pdf).

In Fig.~\ref{fig:hysteresis}, we show the final steady state of the exact same simulation as in Fig.~\ref{fig:no_hysteresis} except that we have slightly increased the feeder length, $L=0.50$.  After restoration of $v_0=1.0$, phase front (blue curve) still forms and propagates into the feeder, however, it becomes stationary at $z\sim0.22$.  The lack of a peak in the real power load density (yellow) shows that there is no motor acceleration in the front implying that it is stationary.  In this case, the reversal of the initial perturbation does not restore the original state and the feeder displays significant hysteresis.  The added reactive power load density $q$ between $z=0.45$ and $0.50$ interacts with the impedance over the entire length of the feeder to create conditions (i.e. $v<v_c^+$) near $z\sim 0.25$ that enable the local motor hysteresis to make the feeder globally hysteretic.  The dynamic version of this simulation is provided in the SI (Movie 3: movie\_hysteresis.pdf).  We note that a large section of motors to the right of the stationary phase front have $v_c^-<v<v_c^+$  and therefore also have a stable normal state (see Figs.~\ref{fig:hysteresis-single} and \ref{fig:landau}).  Additional small perturbations could result in local transitions to the normal state and subsequent global recovery.

In between $L=0.45$ and $0.50$ is a critical length $L_c$ where the hysteresis first appears.  Feeders operating with $L>L_c$ can be called ``dangerous'' because they are operating normally until a voltage fault occurs. Post fault, a fraction of the feeder does not recover to normal operation and this fraction of induction motors remains stalled, i.e. the feeder has just undergone a FIDVR event.  Over a period of one to several minutes, the stalled motors will get disconnected by tripping of thermal protection systems, however, this uncontrolled recovery may also lead to overvoltages that are equally troublesome.

\subsection{Special Features of the Phase Transition}
\label{subsec:special}

Although the dynamical behavior described above is reminiscent of many of phase transitions in physics, it also displays some very unique features, e.g. its lack of explicit spatial locality and the instantaneous nature of the global voltage adjustment.  In the Ginzburg-Landau (GL) theory of first-order phase transitions, we also have a potential with two minima, however, the spatio-temporal dynamics of the phase transition are different.  In contrast to the spatio-temporal dynamics of the current problem (as given in Eq.~(\ref{eq:energy_dynamics})), GL dynamics are typically driven by a dispersion term, i.e. by adding a term such as  $\partial_z^2\omega$ to the right hand side of Eq.~(\ref{eq:energy_dynamics}) which effectively couples the order parameter $\omega$ at different spatial locations. The resulting GL dynamics is a phase transition front with a shape and speed defined by the {\it local} balance (i.e. within the front) of the added dispersion term against the existing nonlinearity (rhs of Eq.~(\ref{eq:energy_dynamics})) and dynamics (lhs of Eq.~(\ref{eq:energy_dynamics})). In contrast, the transition front dynamics of the present problem is driven by the globally-superimposed spatial inhomogeneity of the voltage profile $v(z)$.  During short periods of time when there are not abrupt {\it global} changes, the voltage profile remains relatively frozen and the motors respond locally (Eqs.~(\ref{eq:energy_dynamics},\ref{energy})) to the mismatch between their current rotational state $\omega$ and minimum $U(\omega)$ as determined by the local voltage $v(z)$.  These adjustments occur most rapidly in the vicinity of the phase front where the state mismatch is the largest. The adjustment of $v(z)$ to the evolving motor loads is instantaneous (Eqs.~(\ref{p_cont}-\ref{v_eq},\ref{p-motor-cont}-\ref{bc})), but the evolution is actually temporally slow because the motor states are only adjusting in the small region of the phase transition front.

Another point of comparison is the so-called Stefan problem (see \cite{72Rub,Stefan_wiki} and references therein) describing a phase transition driven by heat released at the interface between the two phases. In its one dimensional formulation, the Stefan problem considers two sub-domains with their outer boundaries (the boundaries away from the phase front) maintained at different conditions, e.g. one at a constant temperature flux and another at different temperature. Within each of the sub-domains temperature plays the role of the order parameter, and it obeys simple thermal diffusion (possibly with different diffusion coefficients in the two sub-domains). The sharp phase-phase interface is subject to a boundary condition that relates its speed to the temperature at the interface.  If the interface progresses, heat is released locally and it is then transported via diffusion.  Different versions of the problem show many interesting behaviors.  In a semi-infinite domain, self-similar continually slowing fronts emerge.  In finite domains, the fronts may become stationary. This behavior is similar to what we observe in the present problem -- the analogy with the Stefan problem is the emergence of a global solution with an inhomogeneous order parameter profile and sharp interfacial boundary.  However, the physics and interplay of the mechanisms that creates the behavior in the two problems are significantly different -- thermal diffusion vs heat release in the Stefan problem, and local frequency transformations vs voltage profile and rearrangement. Moreover,  the latter point also emphasizes the difference -- voltage profile plays the role of the heat injection but it does it globally along the feeder, also changing instantaneously, i.e. voltage rearrangement takes place with an infinite speed (electro-dynamic effects are instantaneous in our electro-mechanical model)  while the heat injection condition is modified gracefully in the Stefan case, as the heat propagates along the domain with a finite speed controlled by thermal diffusion.

\section{Simulating and Explaining Fault Induced Delayed Voltage Recovery (FIDVR)}
\label{sec:FIDVR}

In the previous Sections, we built up an understanding of the different types of dynamical behavior of an induction motor-loaded distribution feeder and the different final states that may result.  In this Section, we discuss in more detail the processes specific to FIDVR.  Although we discussed the anatomy of a FIDVR event in the introduction, we repeat it here to motivate the following study of FIDVR dynamics.  Prior to the transmission fault, the voltages $v_0$ at the head of all the distribution feeders extending from the substation served by the transmission line are 1.0, and all the feeders are in the normal phase, i.e. all the induction motors are in the normal state. During a transmission fault, the large fault currents in the transmission lines locally depress the transmission voltage which in turn depresses $v_0$.  $v_0$ remains depressed below 1.0 until the transmission fault has cleared, i.e. automatic protection circuit breakers have opened to  de-energize the faulted line to extinguish the ionized air supporting the fault current and then reclosed to re-energize the transmission line. During the fault-clearing process, one transmission line (out of typically two or more) serving the substation is briefly removed from service so which also tends to depress $v_0$.  It is during this period of $v_0$ depression that the induction motors on the distribution feeders undergo ``collapse dynamics'' and may stall.  If the transmission fault is cleared normally, the voltage at the the substation returns to near normal levels after the circuit breakers reclose.  If there is significant motor stalling, $v_0$ may not fully recover to 1.0 , but we approximate the post-fault voltage as $v_0=1.0$.  The induction motor-loaded feeders then undergo ``recovery dynamics'' which evolve to a final steady state.

In Section \ref{subsec:fault}, we first discuss the collapse dynamics during a period of the fault when $v_0$ is depressed.  This is followed by Section \ref{subsec:recovery} where we discuss the recovery dynamics (or partial recovery) after $v_0$ is restored. Finally, in Section \ref{subsec:dyn_trans}, we analyze the two periods in combination asking the question: under which conditions does a fault result in a FIDVR event?

\subsection{Dynamics During a Fault--Collapse to Stalled Sates}
\label{subsec:fault}

\begin{figure*}
    \centering
    \subfigure[]{\includegraphics[width=0.24\textwidth]{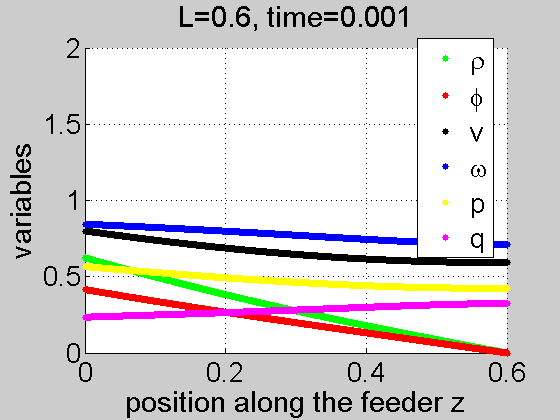}}
    \subfigure[]{\includegraphics[width=0.24\textwidth]{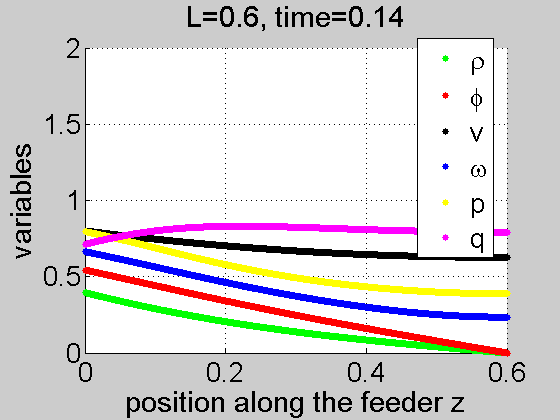}}
    \subfigure[]{\includegraphics[width=0.24\textwidth]{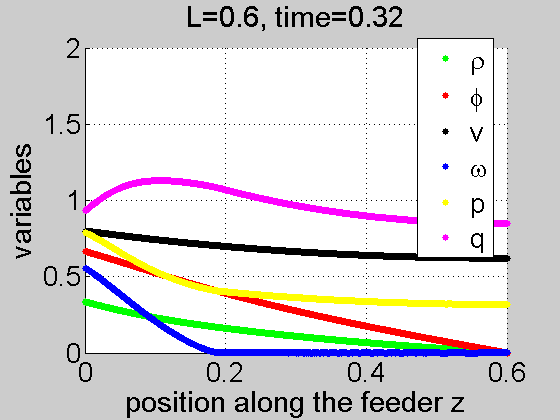}}
    \subfigure[]{\includegraphics[width=0.24\textwidth]{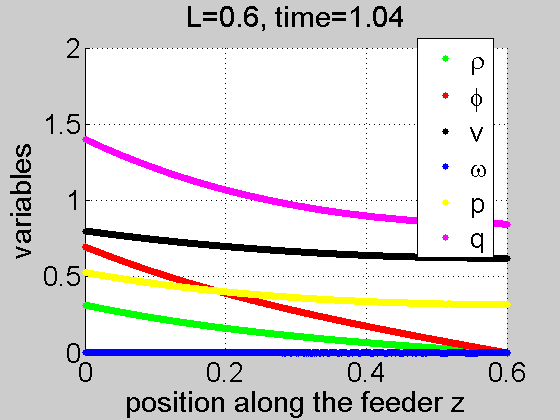}}
\caption{Sequence of snapshots illustrating the collapse dynamics during a large voltage fault with  $v_0<v_c^-$ and $L=0.6>L_c$. Snapshot (a) shows the situation just after the application of the fault -- although the $v(z)<v_c^-$, the motors continue to rotate with $\omega/\omega_0\sim 1$ because of their inertia.  Snapshot (b) corresponds to a short time after the application of the fault when the motors are just starting to decelerate.  The more remote motors experience smaller $v$ and steeper $U(\omega)$. Their faster deceleration results in smaller $\omega/\omega_0$ at these remote locations. The collapse is reinforced by an increase in $q$ (pink) as the motors reach lower $\omega$. Snapshot (c) is taken later in the process: the end of the line is completely stalled, the wave of deceleration starts to propagate backwards, from the tail to the head. Snapshot (d) shows the final phase: all the motors are stalled. Note a significant increase in the reactive power drawn by the stalled feeder. See Movie 1: movie\_large\_fault.pdf in SI for a dynamic version of this process.}
    \label{fig:large_fault}
\end{figure*}

If the voltage drop during the fault is large enough, i.e. if $v_0<v_c^-$ (see Figs.~\ref{fig:torques},\ref{fig:hysteresis-single},\ref{fig:landau}), all the motors will rapidly decelerate to $\omega/\omega_0 \approx 0$ and the entire feeder will eventually end up in the stalled state.  Figure~\ref{fig:large_fault} (and movie 1:movie\_large\_fault.pdf in SI) illustrates the details of this process which are consistent with our interpretation in terms of an ``overdamped fall'' down the the $v<v_c^-$ energy landscape of Fig.~\ref{fig:landau} (dark blue curve) governed by Eq.~(\ref{eq:energy_dynamics}).  We note that the long-range spatial coupling forces the motors farther down the feeder to collapse to $\omega/\omega_0\sim 0 $ faster and reinforces the collapse after is starts. Specifically, the cumulative effect of the induction motors loads reduces the voltage at the remote locations of the feeder (black curve) resulting in a steeper local potential $U(\omega)$ (dark blue curve in Fig.~\ref{fig:landau}).  Eq.~(\ref{eq:energy_dynamics}) shows that these remote motors stall sooner, which is borne out in Fig.~\ref{fig:large_fault}c (blue curve).  Additionally, as the remote motors decelerate, their local reactive load density $q$ (pink) increases which increases $\phi$ (red) further lowering $v(z)$ (black) and increasing the slope of the local potential $U(\omega)$.  The result is an even faster collapse into a pure stalled phase.  Moreover, we observe that the more severe the voltage drop or the lower the motor rotational inertia $\mu$, the earlier these motors will be stalled.

\begin{figure*}
    \centering
    \subfigure[]{\includegraphics[width=0.24\textwidth]{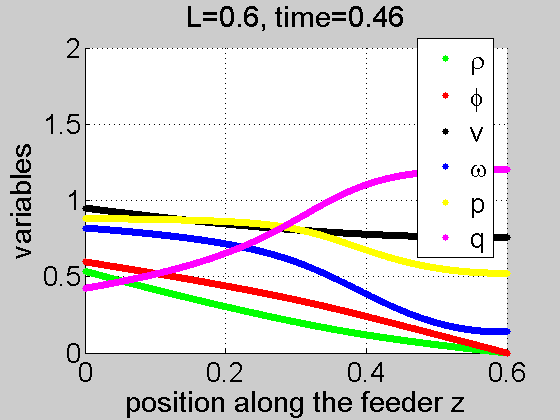}}
    \subfigure[]{\includegraphics[width=0.24\textwidth]{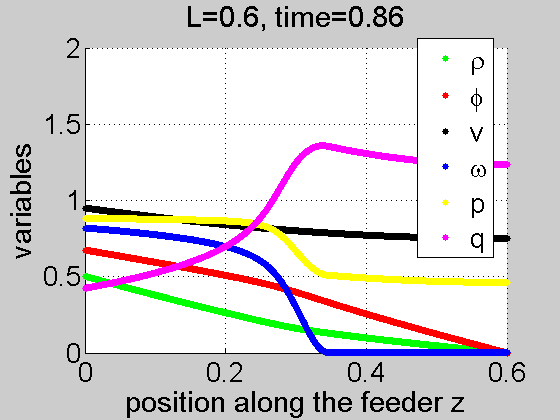}}
    \subfigure[]{\includegraphics[width=0.24\textwidth]{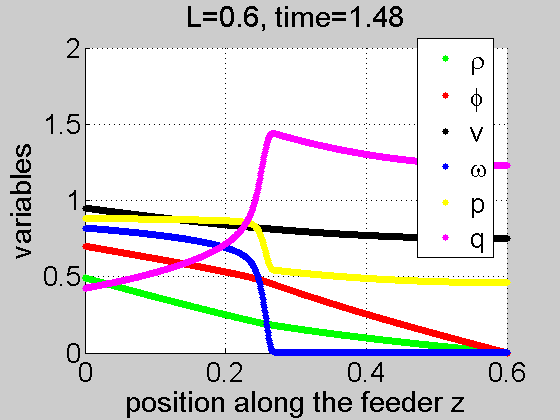}}
    \subfigure[]{\includegraphics[width=0.24\textwidth]{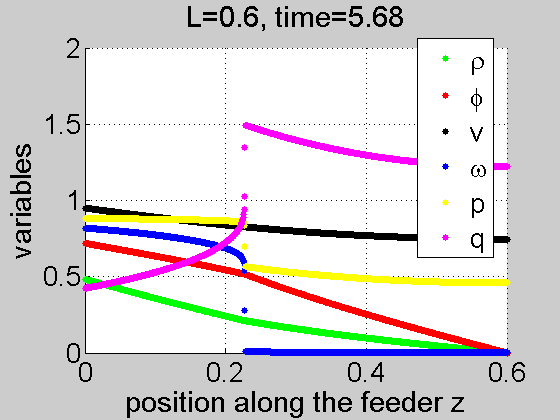}}
    \caption{Collapse dynamics caused by a small voltage fault with $v_0>v_c^-$ and $L=0.6>L_c$.  The collapse proceeds slower than in Fig.~\ref{fig:large_fault} because the higher voltages result in shallower $U(\omega)$.  Snapshot (a) corresponds to a short time after the fault as the motors have just begun to decelerate, however, the rise in $q$ (pink) is already reinforcing the collapse.  Snapshot (b) corresponds to later during the fault where the end of the line is stalled and the deceleration front is starting to feel the influence of the boundary condition at $z=0$.  Snapshot (c) the deceleration front has become nearly stationary as the long-range interactions can no longer overcome the boundary condition at $z=0$.  Snapshot (d) show the feeder in the stabilized, partially-stalled steady state.See Movie 2: movie\_small\_fault.pdf in SI for a dynamic version of this process.}
    \label{fig:small_fault}
\end{figure*}

Next we consider a less severe fault where $v_c^-<v_0<1.0$ (see Fig.~\ref{fig:small_fault} and the full movie version, Movie 2: movie\_small\_fault.pdf in the SI).  Although $v_0>v_c^-$, the immediate post-fault voltage drops along the feeder to a point $z_0$ where $v(z_0)<v_c^-$.  If the voltage profile was subsequently frozen in time, we would expect the motors with $z>z_0$ to behave very much as in Fig.~\ref{fig:large_fault}, i.e. a decleration to $\omega/\omega_0\sim 0$, although somewhat slower in the previous case because the local $v(z)$ is slightly higher.  In Fig.~\ref{fig:small_fault}b, we do observe this initial behavior, however, the local reactive power loading $q$ (pink) again increases as the motors stall increasing $\phi$ (red) and lowering the local $v(z)$.  Via this spatial coupling, the feeder power flows reinforce the collapsing wave allowing it to propagate to $z<z_0$.

Although the reinforcement process can be significant, it cannot overcome the the boundary condition at $z=0$ that maintains $v_0>v_c^-$.  As shown in Fig.~\ref{fig:small_fault}d, the phase front finally stops at location $0<z<z_0$ where the local voltage finally stabilizes at $v(z)=v_c^-$.  As this occurs, local motor accelerations cease and the phase front in $\omega$, $p$, and $q$ steepen creating a sharp demarcation between the two pure phases of normal and stalled states.  One common and possibly universal feature in the two scenarios discussed above is emergence of a ``collapse''  transient which can be interpreted as a quasi-stationary phase transition front of a slowly evolving ``soliton'' shape propagating with the speed and shape controlled by instantaneous  voltage at the point of the front which defines the slope of the   energy landscape $U(\omega)$ in Eq.~(\ref{energy}) and Fig.~\ref{fig:landau}.

\setcounter{subfigure}{0}

\subsection{Post-Fault Dynamics--Recovery from Stalled (or Partially Stalled) States}
\label{subsec:recovery}

The severity of the $v_0$ depression during the fault determines the final state that will be reached if the collapse dynamics are allowed to evolve to steady state.  The transmission fault may be cleared ($v_0$ returns to 1.0) before this steady state is reached, and we explore the dependence of the recovery dynamics on the fault time duration in Section~\ref{subsec:dyn_trans}.  For simplicity of the present discussion, here we assume we are starting from a post-collapse steady state.  Whether the feeder starts in a partially or fully-stalled post-collapse steady state, the perturbation of $v_0$ returning abruptly to 1.0 has the chance of restoring the feeder to the fully normal phase.

\begin{figure*}
    \centering
    \subfigure[]{\includegraphics[width=0.24\textwidth]{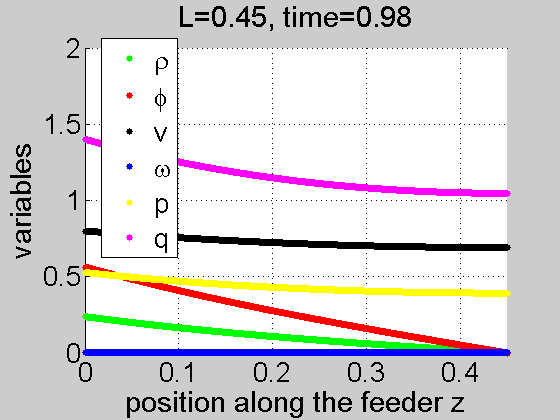}}
    \subfigure[]{\includegraphics[width=0.24\textwidth]{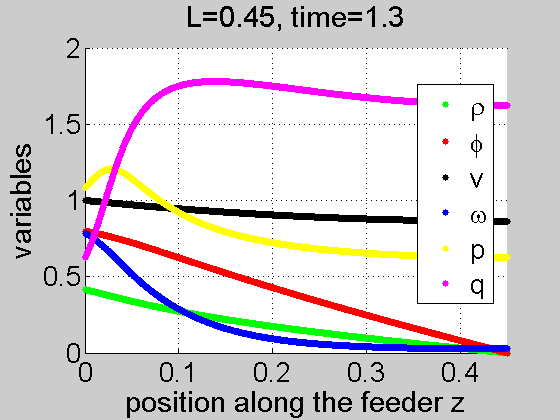}}
    \subfigure[]{\includegraphics[width=0.24\textwidth]{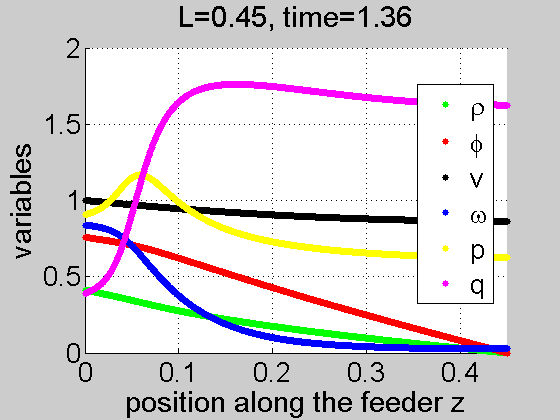}}
    \subfigure[]{\includegraphics[width=0.24\textwidth]{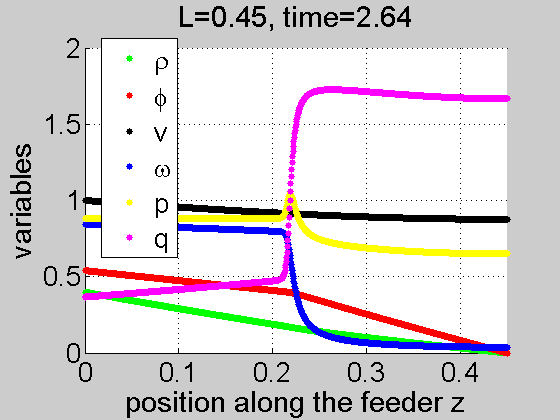}}
    \subfigure[]{\includegraphics[width=0.24\textwidth]{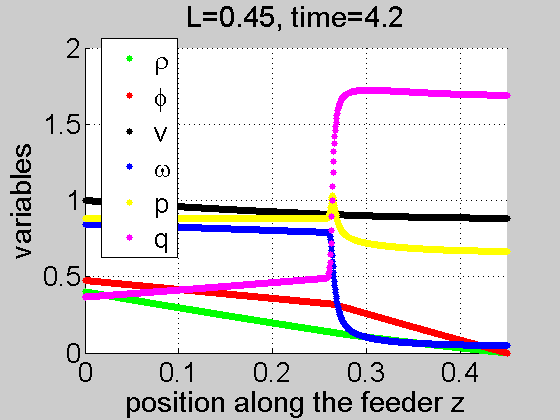}}
    \subfigure[]{\includegraphics[width=0.24\textwidth]{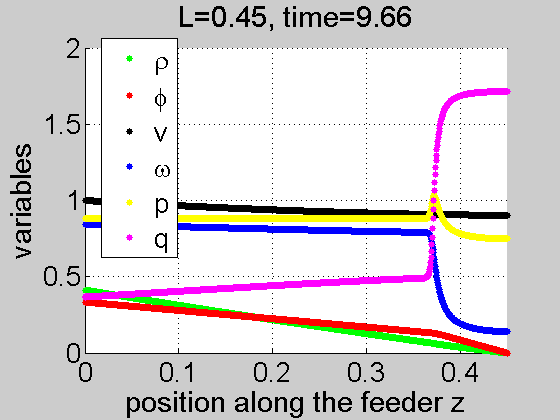}}
    \subfigure[]{\includegraphics[width=0.24\textwidth]{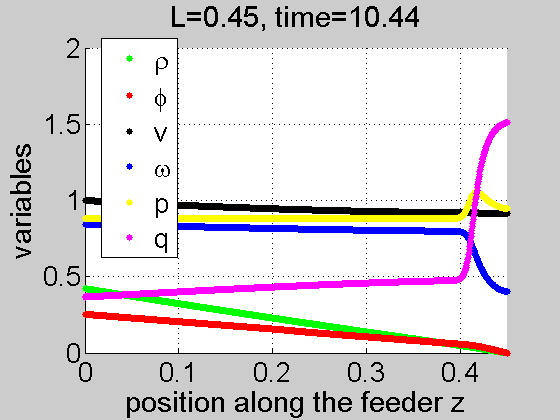}}
    \subfigure[]{\includegraphics[width=0.24\textwidth]{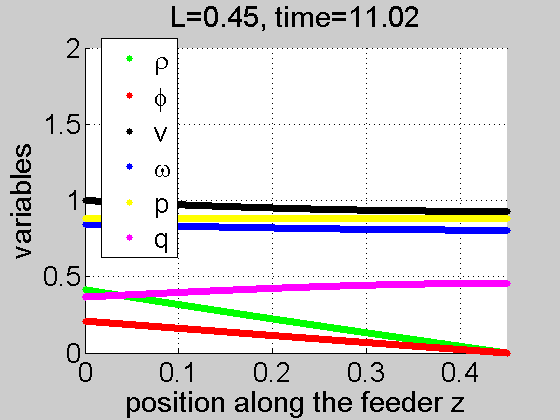}}
    \caption{Eight snapshots of the evolution of a feeder from a steady-state, fully-stalled phase to the fully-normal phase following the restoration of $v_0=1.0$.  (a) The final fully-stalled steady state reached after simulating the feeder for a $v_0=0.8$ fault voltage depression.   (b) $v_0$ is restored to 1.0 and the motors at the head of the feeder start to accelerate. (c) A recovery front is built which starts to propagate into the feeder. (d) The front continues to propagate at roughly constant speed and also showing a universal ``soliton"-like shape. (e) The front continues to propagate with roughly the same speed and a universal ``soliton"-like shape. (f) Completion of the recovery process where interaction with the end of the feeder accelerates the recovery front. (g) The recovery front still propagates while the shape of the front starts to change because fo interactions with the end of the feeder.  (h) End of the recovery process: the entire feeder is back to the normal state. See Movie 4: movie\_recovery.pdf in SI for dynamical illustration.}
    \label{fig:beg_soliton_end}
\end{figure*}

We have performed a range of simulations starting from both partially and fully-stalled steady states resulting from simulation of a depressed of $v_0$.  We then restore $v_0=1.0$ and monitor the recovery dynamics.  One such case for a fully-stalled initial condition that fully recovers is shown in Fig.~\ref{fig:beg_soliton_end}.  When restoring $v_0=1.0$ is sufficient to drive a full recovery (as in Fig.~\ref{fig:beg_soliton_end}), we observe a rather rich dynamics which can be split, roughly, into the following three stages:
\begin{itemize}
  \item Figures~\ref{fig:beg_soliton_end}a$\rightarrow$b: The abrupt change of $v_0$ instantaneously creates a smooth voltage profile dependent upon the instantaneous adjustment of real $p$ and reactive $q$ load densities to the higher voltage, but at their motor's stalled rotation speed.  Motors with $v(z)>v_c^+$ start to accelerate, and the phase transition front is set up in the vicinity of the feeder head.

  \item Figures~\ref{fig:beg_soliton_end}c$\rightarrow$e: The recovery front matures and sharpens while propagating into the feeder.  The narrow recovery front is always located near to $v(z)=v_c^+$ where the fast dynamics of motor acceleration and state transition occur.  These fast transitions are accompanied by jumps in $p$ and $q$ which act in a non-local manner to push the location of $v(z)=v_c^+$ to larger $z$ thus driving the transition front farther into the feeder.  The propagation, although driven by the fast dynamics of motor state transitions, comprises a slow dynamics because the state transitions occur in a phase front that is narrow compared to the feeder length.  Also part of the phase front is a peak in $p$ above the steady-state values on either side of the front.  This peak is required to supply the power to accelerate the motors from a stalled to a normal state.  If the feeder is long enough, the phase front appears to reach a time-invariant shape that propagates much like a soliton.

  \item Figures~\ref{fig:beg_soliton_end}f$\rightarrow$h: The slow dynamics of the recovery approaches to within about one or two phase front widths of the end of feeder.  With fewer motors to accelerate, the voltage adjusts faster and the front propagation speed increases until is has consumed the entire feeder and the feeder reaches a uniform normal phase.
  See Fig. \ref{fig:beg_soliton_end}f-h for illustration.
\end{itemize}
See Movie 4: movie\_recovery.pdf in SI for the full movie of the recovery phenomenon.

\setcounter{subfigure}{0}

Perhaps the most remarkable feature of the recovery process is formation, at the intermediate stage, of a ``soliton"~-- a quasi-stationary shape moving with roughly constant speed from the head to the tail. The emergence of the ``soliton'' is due to the long-range interaction of the fast but local changes in $p$ in $q$ in the phase front to the slower but global changes  in voltage $v(z)$.

As we will see in the following Section, the feeder does not always fully recover simply because $v_0$ is restored to 1.0.  In these cases, we have explored the effects of temporarily (or permanently) raising $v_0$ above 1.0 and found this approach to be quite effective in restoring feeders that would have otherwise remained stalled.  However, such intelligent control action requires reliable detection of the entry into a FIDVR event and fast switching and/or device control to increase $v_0$ to a sufficient level.  We postpone full analysis of such a control to future work.



\subsection{Dynamic Transition: Will a Feeder Enter a FIDVR State?}
\label{subsec:dyn_trans}

\begin{figure}
	\centering
	 \includegraphics[width=0.48\textwidth]{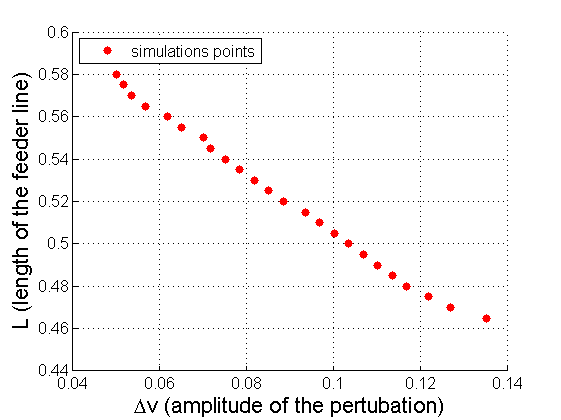}
	\caption{Phase diagram of the dynamic transition from the normal phase to a partially stalled phase. All the points above the curve lead to a partially stalled phase, all the points under the curve result in a normal state. For $L\leq L_c\simeq 0.46$, there is no hysteresis and the system always ends in a fully running phase.}
	\label{fig:L_vs_Deltav}
\end{figure}

\begin{figure}
	\centering
	 \includegraphics[width=0.48\textwidth]{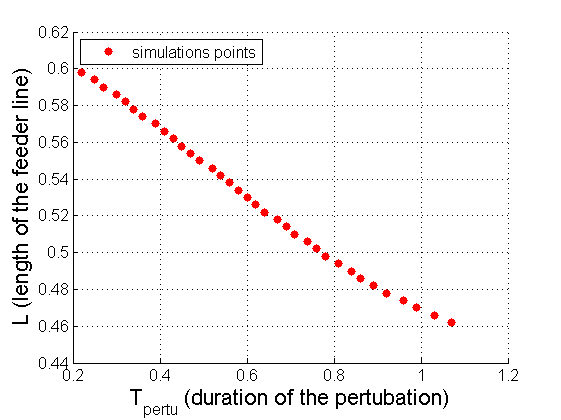}
	\caption{Phase diagram of the dynamic transition from the normal phase to a partially stalled phase. All the points above the curve lead to a partially stalled phase, all the points under the curve result in a normal phase. For $L\leq L_c\simeq 0.46$, there is no hysteresis and the system always ends in a fully running phase.}
	\label{fig:L_vs_Tpertu}
\end{figure}

\begin{figure}
	\centering
	 \includegraphics[width=0.48\textwidth]{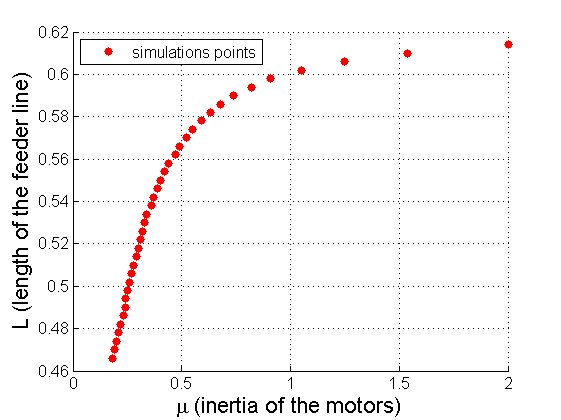}
	\caption{Phase diagram of the dynamic transition from the normal phase to a partially stalled phase. All the points above the curve lead to a partially stalled phase, all the points under the curve result in a normal phase. For $L\leq L_c\simeq 0.46$, there is no hysteresis and the system always ends in a fully running phase.}
	\label{fig:L_vs_mu}
\end{figure}

\begin{figure}
	\centering
	 \includegraphics[width=0.48\textwidth]{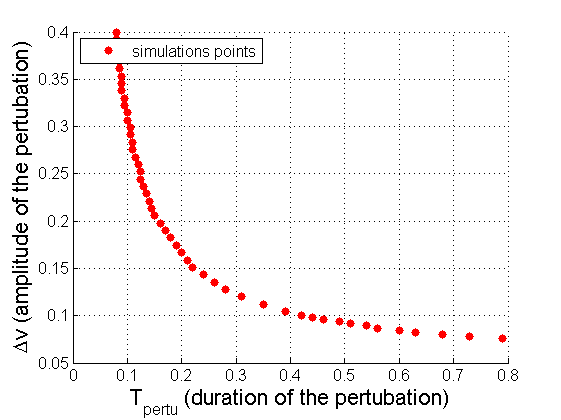}
	\caption{Phase diagram of the dynamic transition from the normal phase to a partially stalled phase. All the points above the curve lead to a partially stalled phase, all the points under the curve result in a normal phase.}
	\label{fig:Deltav_vs_Tpertu}
\end{figure}

\begin{figure}
	\centering
	\includegraphics[width=0.48\textwidth]{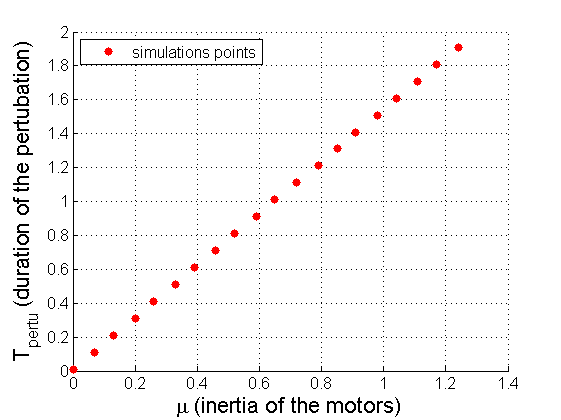}
	\caption{Phase diagram of the dynamic transition from the normal state to a partially stalled state. All points under the curve lead to a partially stalled regime, all points above the curve result in the normal final state.}
	\label{fig:Tpertu_vs_mu}
\end{figure}

\begin{figure}
	\centering
	 \includegraphics[width=0.48\textwidth]{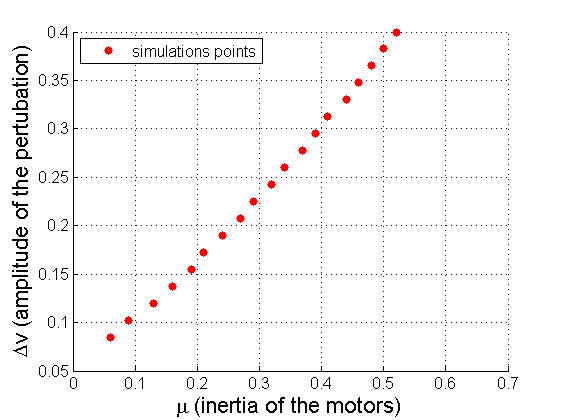}
	\caption{Phase diagram of the dynamic transition from the normal state to a partially stalled phase. All the points above the curve lead to a partially stalled phase, all the points under the curve result in a normal phase.}
	\label{fig:Deltav_vs_mu}
\end{figure}

Our simulation results suggest that the normal phase, the fully stalled phase, and any of the partially stalled phases (with the feeder split in the normal head and stalled tail) can be the final stationary and stable point of a dynamical evolution. In this Section, we explore the properties of the final state as the properties of the feeder (length $L$ and motor inertia $\mu$) and the fault (magnitude of voltage depression $\Delta v$ and duration $T_{pertu}$) are varied. Our goal is to develop an initial understanding of the ``non-equilibrium phase diagram'' that controls whether or not the feeder recovers to a fully normal phase.  In the test discussed below, we consider a feeder with $L>L_c$ because those with $L<L_c$ are known to always recover to a fully normal phase no matter the size or duration of the perturbation.  In our study of the phase diagram, we dissect the four-parameter space, $(L, \mu, \Delta v, T_{pertu})$ in six different ways:
\begin{enumerate}

\item fixing $\mu$ and $T_{pertu}$ and exploring the $(L,\Delta v)$ subspace, see Fig.~\ref{fig:L_vs_Deltav};
\item fixing $\mu$ and $\Delta v$ and exploring the $(L,T_{pertu})$ subspace, see Fig.~\ref{fig:L_vs_Tpertu};
\item fixing $\Delta v$ and $T_{pertu}$ and exploring the $(L,\mu)$ subspace, see Fig.~\ref{fig:L_vs_mu};
\item fixing $L$ and $\mu$ and exploring the $(T_{pertu}, \Delta v)$ subspace, see Fig.~\ref{fig:Deltav_vs_Tpertu};
\item fixing the values of $L>L_c$ and (sufficiently large) $\Delta_v$ and  exploring the $(\mu, T_{pertu})$ subspace, see Fig.~\ref{fig:Tpertu_vs_mu};
    \item fixing $L$ and $T_{pertu}$ and exploring the $(\mu,\Delta v)$ subspace, see Fig.~\ref{fig:Deltav_vs_mu}.
\end{enumerate}
Starting as usual with the feeder in a normal state with $v_0=1$, we apply a $v_0$ depression of magnitude $\Delta v$ and duration $T_{pertu}$ with subsequent recovery to $v_0=1$ and then integrate the dynamics until the feeder reaches a steady state.  Unless all of the motors on the feeder recover to a normal state, the feeder final state is classified as partially stalled.  In all of the subsequent Figures, the filled red circles indicate this boundary between a fully normal phase feeder and a partially-stalled feeder.

From our studies of the subspaces and the following Figures, we can make the following conclusions:
\begin{itemize}
  \item From Figs.~\ref{fig:L_vs_Deltav}-\ref{fig:L_vs_mu}, it is apparent that the feeder becomes less resilient to perturbations as it grows in length and that there is an upper feeder length $L^*\approx 0.63$ that requires an infinitesimal perturbation to force it into a partially-stalled phase. In fact, if the feeder is too long (i.e. $L>L^*$), the normal phase is no longer stable ($v(L)\leq v_c^-$) and the feeder always has a partially stalled phase.  Therefore, for the remainder of this study, feeder with $L_c<L<L^*$ are of interest because feeders with $L$ outside this range are either robust to all perturbations or always unstable to a partially-stalled phase.
      \item From Fig.~\ref{fig:Deltav_vs_Tpertu}, it is apparent that there is value of $\Delta v\approx 0.05$ such that the feeder will never stall no matter how long the perturbation is applied.  At $\Delta v =0.05$, $v_0=0.95$ and the static voltage drop at these reduced voltages would make $v(L)\approx 0.83=v_c^-$.  From this interpretation, this lower bound on the $\Delta v$ for extremely long $T_{pertu}$ can be computed from the static power flow equations by looking for the $v_0$ that forces $v(L)=v_c^-$.
  \item  The two previous conditions can be computed from static considerations. The interesting dynamics behind the transition to a partially-stalled state is then expressed in Figs.~\ref{fig:Tpertu_vs_mu} and \ref{fig:Deltav_vs_mu} which can be understood by considering the decline of the motor rotational frequency that occurs at the far end of the feeder during the time $T_{pertu}$ of the fault.  We can crudely approximate this decline by the product of fault duration and the rate of frequency decline immediately after the application of the fault, i.e. $\Delta (\omega/\omega_0)|_{t=T_{pertu}} \approx   -(2P/v_0\omega_0^2)  (T_{pertu} \Delta v/\mu)$ where we have used Eq.~\ref{torque-cont} and a linear expansion of Eq.~\ref{p-motor-cont}.  It is reasonable to expect that the boundary between a feeder that fully recovers and one that is partially stalled would be expressed by $\Delta (\omega/\omega_0)|_{t=T_{pertu}} \approx$ constant.  If such a relationship is found to hold, it would imply $(T_{pertu} \Delta v/\mu)\approx$ constant, which within scope of our parametric study, is in rough agreement with the results in  Figs.~\ref{fig:Tpertu_vs_mu} and \ref{fig:Deltav_vs_mu}, especially if we account for the minimum required value of $\Delta v$ discussed immediately above.
\end{itemize}
The approximate analysis above provides a good qualitative understanding of the fault and feeder parameters that lead to a feeder with a partially-stalled phase.  However, more rigorous analytical analysis of Eqs.~\ref{p_cont}-\ref{bc} is required to put these conclusions on firm footing.

\begin{figure*}
    \centering
    \subfigure[$T_{pertu}=0.221$ and $\Delta v =0.15$]{\includegraphics[width=0.24\textwidth]{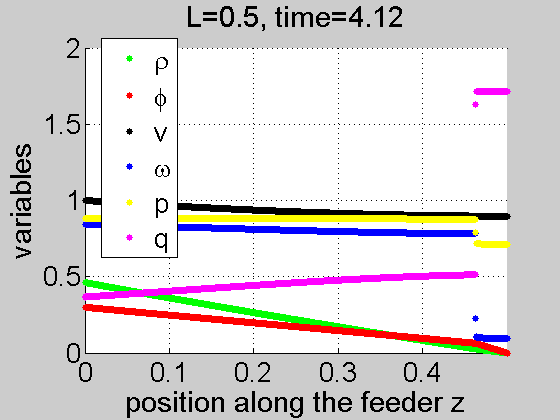}}
    \subfigure[$T_{pertu}=0.23$ and $\Delta v =0.15$]
	{\includegraphics[width=0.24\textwidth]{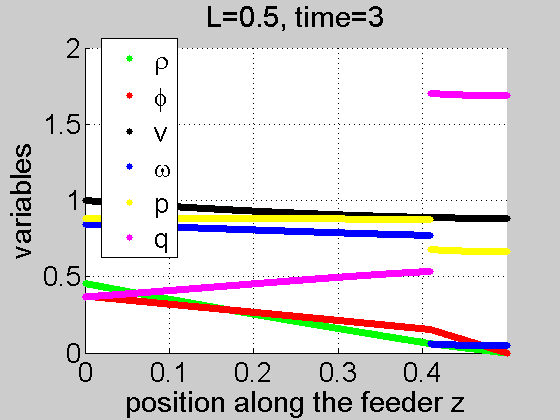}}
    \subfigure[$T_{pertu}=0.3$ and $\Delta v =0.15$]{\includegraphics[width=0.24\textwidth]{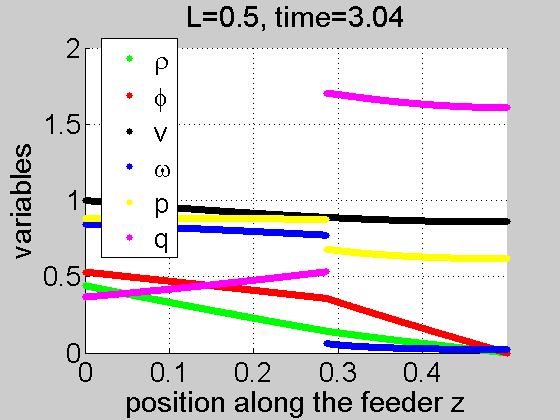}}
    \subfigure[$T_{pertu}=0.5$ and $\Delta v =0.15$]{\includegraphics[width=0.24\textwidth]{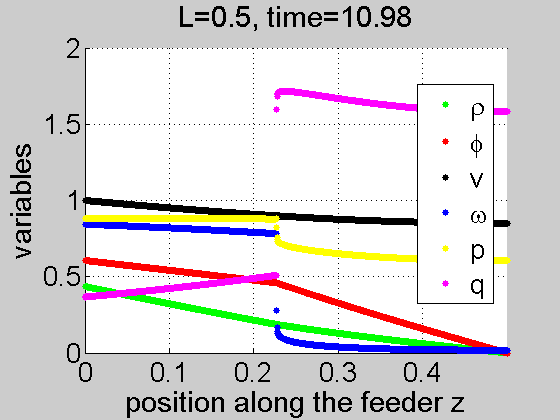}}
    \caption{Different final states can be achieved by manipulating the duration of the perturbation $(T_{pertu}$),  while its amplitude $\Delta v$ remains constant. From left to right, the fault is longer and longer and the number of stalled motors increases. See file movie\_different\_states.pdf in SI for five movies illustrating the different phases that can be reached and the dynamics that lead to the phases.}
    \label{fig:different_states}
\end{figure*}

Beyond just determining the boundary between a fully-normal and partially-stalled feeder final states, the characteristics of the voltage fault also determine the number of motors that will be stalled after the fault is cleared.  The more severe the fault (in duration $T_{pertu}$ or in amplitude $\Delta v$), the more motors will get stalled, up to a point. If the fault is severe enough ($\Delta v \geq \Delta v^*$ or $T_{pertu} \geq T_{pertu}^*$), the number of stalled motors will not increase any more.  For example, the maximum number of stalled motors is reached in Fig.~\ref{fig:different_states}d for a given set of parameters of the feeder line, and an even more severe fault will end up in this same state after it is cleared.

On the other hand, the less severe the fault the less the number of stalled motors, and the system can reach a continuous set of stable partially-stalled phases between the maximum number of stalled motors and none of the motors stalled. Fig.~\ref{fig:different_states} shows examples of the different phases the system can reach, and the file movie\_different\_states.pdf in SI illustrates the dynamics leading to these final states.

\setcounter{subfigure}{0}

\section{Discussions \& Path Forward}
\label{sec:disc}

In this manuscript,  we have modeled, simulated, analyzed and explained how the electro-mechanical dynamics of induction motor-loaded distribution feeders can lead to Fault-Induced Delayed Voltage Recovery (FIDVR), an effect observed recently in power distribution feeders. We approached these dynamics and the FIDVR events from the stand-point of physics -- explaining and interpreting them as an instance of a broader class of nonlinear electro-mechanical phenomena in power distribution systems. The main ideas and questions discussed in this work can be summarized as follows:
\begin{itemize}
\item The 1+1 (space+time) continuous model, introduced in \cite{11CBTCL} and developed here, offers a computationally efficient framework for analysis of nonlinear and dynamical phenomena in distribution feeders, and crucially, the PDE model enables analogies with other known spatio-temporal dynamical systems and solutions to help build intuition about the dynamical behavior of the electrical grid's  electro-mechanical dynamics.  
\item By coupling a spatially local model of an individual motor -- that describes its nonlinear, bi-stable and hysteretic switching between normal and the stalled states~-- to a continuum version of the electrical power flow equations, we are able to explain the coexistence of a spatially-extended normal phase with a stalled phase (motors at the head portion of the feeder are in the normal state, while motors in the tail portion of the feeder are stalled).
\item The emergence of the multiple spatially-extended states is interpreted in terms of a first-order phase transition where the distribution of the motor rotational frequency along the feeder is the order parameter. The voltage distribution along the feeder plays the role of an external field (degree of freedom) that modifies an effective energy potential for motor frequency. Different (normal, stalled or partially stalled) spatially-extended phases are stabilized and reach a steady state by achieving a global electro-mechanical balance of their voltage and frequency distributions along the feeder.
\item Sufficiently strong perturbations, e.g. sudden drops or rises in voltage at the head of the feeder, lead to transients in the form of propagating phase fronts that separate normal-state motors at the head of the feeder from stalled-state motors in the tail of the feeder.  We analyzed and classified the different types of transients and resulting steady-state phase distributions that emerge after the transients settle for different perturbation strengths and lengths of the feeder.
\item We also experimented with the dynamics of mechanical frequency and voltage phase distributions under more realistic, but more complex, two-step perturbations~-- a voltage drop at the head of the feeder followed shortly by restoration back to its nominal value -- that are expected to simulate the real world perturbations that result in FIDVR events. The dynamics and emerging steady states were explored for different voltage perturbations (depth and duration of the voltage depression) and feeder characteristics (feeder length and inertia of the connected induction motors).
\end{itemize}
Major conclusions drawn from our numerical experiments and analysis are
\begin{itemize}
\item \underline{\bf Hysteresis.} The system is strongly hysteretic: reversing a perturbation does not lead to a simple reversal of the dynamical trajectory.
\item \underline{\bf Recovery Conditions.} When the feeder is short enough or the voltage perturbation is weak enough (small enough in amplitude or short enough duration), the feeder recovers to a fully normal phase following a voltage perturbation thus avoiding a FIDVR event.  Longer feeders or stronger voltage perturbations lead to incomplete recovery and, by modifying the three parameters beyond the recovery threshold, one explores a continuous family of different partially-stalled phases.
\item \underline{\bf Self-Similar Transients.} When a feeder is sufficiently long, recovery transients appear to show universal soliton-like phase fronts with normal phase propagating into the stalled phase with an (approximately) constant speed and time-invariant shape.
\end{itemize}

This manuscript opens up a new line of research into physics-based analysis of transients and phase transitions in distribution feeders. We plan to continue this work focusing on the following generalizations and extensions:
\begin{itemize}
\item We modelled distribution feeders as consisting of identical induction motor loads distributed uniformly along the feeder. In reality, the motors may be different and distributed non-uniformly along the feeder, their distribution and parameters may fluctuate in time, and they are present with many other different types of loads. We will extend our purely deterministic analysis to a probabilistic framework to describe the effects of these forms of disorder and noise to resolve questions such as: what is the probability that the feeder with a given level of disorder will recover after a perturbation of given amplitude and duration and not enter a FIDVR state?
\item As recently shown in \cite{12WTC}, controls associated with distributed generation, e.g. inverters coupled to distributed photovoltaic (PV) generation, can also be incorporated into the spatially continuous modelling framework (ODE framework).  The static model of \cite{12WTC} suggests that a feeder with a sufficiently large penetration of the PV generation may show a rather complicated bifurcation diagram with the emergence of multiple low-voltage solutions that are related to the low-voltage, stalled solutions discussed in this manuscript.  Much like the effects discussed here, these inverter-driven low-voltage states are not the result of the behavior of a single inverter, but rather result from the collective action of many inverters and their and interaction with the nonlinearity and nonlocal behavior of the power flow equations.  To address these dynamical problems in a more comprehensive manner, we will extend dynamical description of this manuscript to the case of distributed generation, and more generally, to the case of a feeder containing a mixed portfolio of distributed generation and different types of loads, including induction motors.

\item In our first physics-based paper on the subject of electro-mechanical dynamical transients in distribution feeders, we relied mainly on numerical experiments. However, the problem formulation may allow asymptotic theoretical analysis, in particular:  accurate resolution of the phase transition boundaries, determination of the bifurcation (spinodal) points, propagation of soliton-like phase fronts, and analysis of the tails of the distribution functions that account for the aforementioned effects of noise and disorder.

\item We have focused primarily on describing electro-mechanical dynamics and the perturbations and transients that lead to FIDVR events.  Armed with the comprehensive understanding gained in the process, we are ready to attack the larger question of distribution feeder voltage control.  Specifically, what is the least control effort needed to avoid a FIDVR event following a given type of fault?

\end{itemize}

\appendix

\section{Methods}
\label{sec:methods}

\subsection{Analysis of Static ODE}
To solve numerically  Eqs.~(\ref{p_cont},\ref{q_cont},\ref{v_eq},\ref{torque-cont},\ref{p-motor-cont},\ref{q-motor-cont}) in the stationary (time-independent) case, we proceed as follows:
\begin{itemize}
  \item using the mechanical torque $t=t_0 \left(\frac{\omega}{\omega_0}\right)^{\alpha}$ (i.e $t \in [0, t_0]$), as a scanning parameter, we ``map'' the values of $p$, $q$, $\omega$ and $v$ through equations
    (\ref{torque-cont},\ref{p-motor-cont},\ref{q-motor-cont}));
  \item for all these ``maps'', we choose only one part of the hysteresis curve (either the one when we start from low voltage, or the one from high voltage);
  \item we then solve equations (\ref{p_cont},\ref{q_cont},\ref{v_eq}) using the ``maps'' of the other variables and the matlab \emph{bvp4c} solver.
\end{itemize}
Note that some of the stationary solutions discovered in the results of dynamic exploration were missed by the static analysis detailed above.

\subsection{Dynamic Simulations}
To solve numerically Eqs.~(\ref{p_cont},\ref{q_cont},\ref{v_eq},\ref{torque-cont},\ref{p-motor-cont},\ref{q-motor-cont}), we employ space-time discretization and use an explicit finite differences scheme.
Our dynamical simulations are split in two steps.
\begin{itemize}
	\item During the first sub-step we take $p$ and $q$ fixed (outputs of the previous time step) and solve the feeder-global static Eqs.~(\ref{p_cont},\ref{q_cont},\ref{v_eq}) for $v$, $\phi$ and $\rho$ under conditions of the fixed voltage at the head of the line and zero fluxes at the end of the line (Eq.~(\ref{bc})). This sub-step instantaneously imposes a spatially smooth and globally correlated voltage profile.
	\item Once the global, $v,\rho,\phi$ variables are fully updated, we start updating on the second sub-step the local variables, $\omega,p,q$. For $\omega$, we use its explicit time dependence:
		\begin{align}
			\omega _j ^{t+1} = \omega _j ^t + \frac{\Delta_t}{\mu} \left( \frac{p_j ^t}{\omega_0} - t_0 \left(\frac{\omega _j ^t }{\omega_0} \right)^\alpha 			\right)
		\end{align}
		where $\Delta_t = T/N_t$ is the time step ($N_t$ is the number of steps, and $T$ the time of the simulation). Once $\omega$ is updated, we easily get the updated values of $p$ and $q$ using Eqs.~(\ref{p-motor-cont},\ref{q-motor-cont}).
\end{itemize}

\section{Movies}
\label{sec:movies}

\label{sec:moviesh}
\subsubsection{Movie 1:``movie\_large\_fault.pdf''}

This movie (see \cite{movie1}) shows the dynamical transient following a large drop in $v_0$ to 0.8, i.e. $v_0<v_c^-$.  At $t<0$, the line is in a stable stationary state where all the motors in the normal state.  At $t=0^+$ (the first picture in the movie), $v_0$ suddenly drops from $1$ to $0.8$, and the voltage profile along the line (black curve) immediately responds. In response to the lower $v(z)$, the motors' rotational frequency (blue) decelerates everywhere along the line at a rate dependant on the local value of the voltage -- motors closer to the head of line where the voltage is larger decelerate at a lower rate. Motors at the far end of the line decelerate faster and become stalled first.  The stalling of these remote motors results in an increase in their local reactive load $q$ (pink) and the overall reactive power flow $\phi$ (red) which reinforces the reduction in $v(z)$ creating a normal-to-stalled transition front that propagates from the tail to the head of the feeder.  Eventually, the entire feeder becomes fully stalled.

\subsubsection{Movie 2: ``movie\_small\_fault.pdf''}

This movie (see \cite{movie2}) shows the dynamical transient following a small drop in $v_0$ to 0.95, i.e. $v_0>v_c^-$. The response of the voltage and the mechanical frequency of the motors in the tail portion of the line is similar to that in Movie 1 -- the voltage near the feeder tail is too low and they decelerate with those nearer the tail getting to the stalled $\omega/\omega_0 \sim 0$ state first. The transition to the stalled state expands as it is again reinforced by the increase in reactive loading $q$ (pink) and reactive power flow $\phi$ (red).    However, the phase front slows down, sharpens, and eventually stops propagating as it begins to feel the boundary condition at $z=0$ that does not allow the motors near the head of the feeder to stall.  The result is a half-stalled, mixed phase distribution with motors in the tail part of the line stalled,  while motors near the feeder's head in the normal state. The slowing down of the front is related to the effect of critical slow down typical of  first-order phase transitions as they approach a spinodal point, i.e. as $v(z)$ approaches $v_c^-$.

\subsubsection{Movie 3: ``movie\_hysteresis.pdf'''}

This movie (see \cite{movie3}) shows hysteretic behavior of a feeder with $L>L_c$, i.e. the application of an opposite perturbation does not lead to a reversal of the dynamical trajectory back to the original state.  From $t=0$ to 0.9, the movie is the same as  Movie 1 ``movie\_large\_fault.pdf'', i.e. $v_0$ is reduced to 0.8 at $t=0$ and the feeder and shows a complete collapse into the fully stalled phase.  At $t=0.9$, the perturbation is reversed as $v_0$ is restored to 1.0.  The increased voltage causes the motors at the head of the feeder accelerate and return to the normal state and a propagating phase front is formed (more details about this recovery front can be found in the description of Movie 4 "movie\_recovery.pdf"). The front advances towards the tail until it reaches the point where the voltage has fallen to to $v_c^+$ (voltage above which a stalled motors accelerates). At this point, the phase front can proceed no further and we end up with a partially stalled feeder. The voltage is unable to increase further because of the non-local reinforcement of the voltage drop by the stalled motors beyond the point where $v=v_c^+$. Similarly to what was seen in Movie 2 (and what is generally the case when the system approaches a spinodal point) the dynamics gets slower as the moving front approaches the critical voltage $v_c^+$.

\subsubsection{Movie 4: ``movie\_recovery.pdf''}

This movie (see \cite{movie4}) is identical to Movie 3, except that the length of the feeder line now is significantly shorter with $L<L_c$.  In this case, there is no hysteretic behavior after a voltage fault is cleared --- the feeder recovers completely because the shorter $L$ has diminished the non-local effects of the motors near the tail of the feeder. The recovery process take a long time allowing us to distinguish different phases of the process. First, at $1<t<1.3$ the beginning of the line accelerates quickly; then, at $1.3<t<9.5$, one identifies emergence of the recovery front, which propagates down the feeder with a nearly stationary soliton-like shape.  The recovery front starts to change shape as it approaches and interacts with the boundary condition at the feeder tail.  The absence of motors to accelerate beyond $z=L$  leads to fast recovery of the normal states in the vicinity of the tail for $t>9.5$.

\subsubsection{Movie 5: ``movie\_different\_states.pdf"}

There are in fact five movies (see \cite{movie5}) glued together in one file with a different movie on each page of the .pdf .  Each of the movies shows the dynamics of the feeder line during a voltage fault and after the fault is cleared. Characteristics of the feeder and faults are identical in all the movies, except we increase $T_{pertu}$ from segment to segment. We observe that the number of motors stalled in steady state increases with in $T_{pertu}$ showing that different partially-stalled states can be reached depending on the duration of the fault. However,  we also notice existence of an upper limit to the number of motors that can be stalled, i.e. no matter how severe the fault, half of the line will always recover. The two last movies in the file illustrate this phenomenon, i.e. even though the fourth segment has $T_{pertu}=0.5$ and the fifth segment has $T_{pertu}=1$, they both result in identical steady states with the same number of stalled motors.

\begin{acknowledgments}

We are thankful to David Chassin and Ian Hiskens for very helpful discussions, advice and explanations on the history of the FIDVR-related research and literature. We are also thankful to Igor Kolkolov, Vladimir Lebedev and Konstantin Turitsyn for comments and suggestions. This material is based upon work supported by the National Science Foundation award \# 1128501, EECS Collaborative Research ``Power Grid Spectroscopy" under NMC. The work at LANL was carried out under the auspices of the National Nuclear Security Administration of the U.S. Department of Energy at Los Alamos National Laboratory under Contract No. DE-AC52-06NA25396.

\end{acknowledgments}

\bibliographystyle{unsrt}
\bibliography{Bib/FIDVR,Bib/SmartGrid,Bib/voltage}

\end{document}